\documentclass[12pt,compress]{elsarticle}
\usepackage{amsmath,amssymb,amscd}
\usepackage{color}
\journal{Nuclear Physics B}

\newcommand{\beq}{\begin{equation}}
\newcommand{\eeq}{\end{equation}}
\newcommand{\bea}{\begin{eqnarray}}
\newcommand{\eea}{\end{eqnarray}}
\newcommand{\bse}{\begin{subequations}}
\newcommand{\ese}{\end{subequations}}

\newcommand{\noi}{\noindent}

\newcommand{\nn}{\nonumber}

\newcommand{\ba}{\begin{array}}
\newcommand{\ea}{\end{array}}
\newcommand{\balign}{\begin{align}}
\newcommand{\ealign}{\end{align}}

\newcommand{\ep}{\epsilon}

\newcommand{\mbf}[1]{\mathbf{#1}}

\newcommand{\mbs}[1]{\boldsymbol{#1}}

\newcommand{\mbb}[1]{\mathbb{#1}}

\newcommand{\mc}[1]{\mathcal{#1}}

\newcommand{\mr}[1]{\mathrm{#1}}

\newcommand{\wt}[1]{\widetilde{#1}}

\newcommand{\one}{1\hskip -1mm{\rm l}}

\renewcommand{\d}{\partial}

\renewcommand{\leq}{\leqslant}
\renewcommand{\geq}{\geqslant}

\newcommand{\half}{$\frac{1}{2}~$}
\newcommand{\su}[1]{su($#1$)}

\newcommand\ket[1]{|#1\rangle}

\newcommand{\qbinom}[3]{\genfrac{[}{]}{0pt}{}{\,#1\,}{#2}_{#3}}

\newcommand{\eq}[1]{(\ref{#1})}
\newcommand{\Eq}[1]{Eq.~(\ref{#1})}
\begin{document}

\begin{center}
{\Large \bf \sf Super Rogers-Szeg\"o polynomials associated with  \\   
{$BC_N$} type of Polychronakos spin chains}

\vspace{1.3cm}
{\sf 
B. Basu-Mallick$^1$\footnote{
Corresponding author, Fax:+91-33-2337-4637, 
Telephone:+91-33-2337-5345, 
E-mail address: bireswar.basumallick@saha.ac.in},
and C. Datta$^1$\footnote
{E-mail~address: chitralekha.datta@saha.ac.in},
}

\bigskip

{\em $^1$Theory Division, Saha Institute of Nuclear Physics, HBNI,\\
1/AF Bidhan Nagar, Kolkata 700 064, India}

\end{center}

\bigskip \bigskip
\bigskip \bigskip 

\noi {\bf Abstract}
\vspace {.2 cm}

As is well known, multivariate  Rogers-Szeg\"o polynomials 
are closely connected with the partition functions of the $A_{N-1}$ type of Polychronakos spin chains having long-range interactions. 
Applying the `freezing trick', here we derive the partition
functions for a class of $BC_N$ type of Polychronakos 
spin chains containing supersymmetric
analogues of polarized spin reversal operators 
and subsequently use those partition functions 
to obtain novel multivariate super Rogers-Szeg\"o
(SRS) polynomials depending on four types of variables. 
We construct the generating
functions for such SRS polynomials and  
show that these polynomials can be written as some bilinear 
combinations of the $A_{N-1}$ type of SRS polynomials. 
We also use the above mentioned generating functions to 
derive a set of recursion relations 
for the partition functions of the $BC_N$ type of Polychronakos spin chains involving different numbers of 
lattice sites and internal degrees of freedom.

\vspace{.74 cm}
\noi PACS No. : 02.30.Ik, 05.30.-d, 75.10.Pq, 75.10.Jm 

\vspace {.2 cm}


\newpage

\baselineskip 16 pt

\noi \section{Introduction}
\renewcommand{\theequation}{1.{\arabic{equation}}}
\setcounter{equation}{0}
\medskip
 
Exact solutions of one-dimensional quantum integrable spin chains with long-range interactions and their supersymmetric analogues~\cite{Ha88,Sh88,HHTBP92,BGHP93,Ha93, SS93,
Po93,Fr93,Po94,Hi95npb,BUW99,FG05,BB06,BBS08,BFGR08epl,EFG10, 
BBH10,BPS95,YT96,EFGR05,BFGR08,KK09}
have recently attracted a lot of interest  
due to their connection 
 with surprisingly large number of topics in 
 physics as well as mathematics. Indeed, this type of 
 quantum spin chains have found important
 applications in subjects 
like condensed matter systems obeying generalized exclusion
statistics~\cite{Ha93,KK09,Po06}, 
quantum electric transport phenomena~\cite{BR94,Ca95},
`infinite matrix product states' in conformal field theory~\cite{CS10,NCS11,TNS14,BQ14,TS15,BFG16},
planar ${\mathcal N}=4$ super Yang--Mills theory~\cite{BKS03,BBL09,Bea12},
random matrix theory~\cite{TSA95}, Yangian quantum groups~\cite{BGHP93,Ha93,Hi95npb,BBH10,BK98,Ba99,BBHS07,BE08}
and Rogers-Szeg\"o (RS) polynomials~\cite{Hi95npb,Hi95,Hi97,KKN97,HB00}. 
Haldane and Shastry have pioneered 
the study of quantum integrable spin models with long-range 
interaction, 
by deriving the 
exact spectrum of a spin-\half  
chain with equally spaced lattice sites 
on a circle~\cite{Ha88,Sh88}. All   
spins of this \su{2} symmetric Haldane-Shastry (HS) chain interact with each other
through a pairwise exchange interaction whose strength is inversely proportional 
to the square of their chord distances.
A remarkable feature of this \su{2} HS chain and its \su{m} generalization is that they
exhibit Yangian quantum group symmetry even for finite number of lattice sites. As a result, the  energy eigenvalues
of these spin chains can be expressed in an elegant way 
by using certain sequences of the binary digits `0' and `1',
which are called as `motifs' in the literature \cite{HHTBP92,BGHP93}.
Furthermore, the complete spectra of these 
HS spin chains, including the degeneracy factors of all energy levels, can be reproduced from the energy functions of some one-dimensional classical vertex models \cite{BBH10}. 

A close relation between 
the Hamiltonian of the   
\su{m} HS spin chain and     
that of the \su{m} spin Sutherland model can be established
by using the `freezing trick'~\cite{SS93, Po93}. 
Each particle of the later model, with $m$ number of 
`spin' degrees of freedom, moves on a circle for any finite 
value of the corresponding coupling constant. However, 
in the strong coupling limit, the coordinates of such particles 
`freeze' at the minimum value of the scalar part of the 
potential, which coincides with the equally spaced
lattice points of the HS spin chain.
Moreover, the spin degrees of freedom of these particles essentially
decouples from their coordinate degrees of freedom.
As a result, the \su{m} spin Sutherland model naturally
yields the Hamiltonian of the \su{m} HS model which governs 
the dynamics of the decoupled spin degrees of freedom. 
Such a method of deriving the Hamiltonian of a quantum 
spin chain from that of a spin dynamical model has also 
been applied~\cite{Po93} to the 
case of \su{m} spin Calogero model with confining harmonic potential. This   
led to another quantum integrable spin chain with
long-range interaction, which is 
known in the literature as the \su{m}
Polychronakos or Polychronakos--Frahm (PF) spin chain.
The sites of this  PF spin chain, which   
are inhomogeneously spaced on a line, coincide    
with the zeros of the Hermite polynomial~\cite{Fr93}.
Indeed, the Hamiltonian of the ferromagnetic
\su{m} PF spin chain 
is given by    
\beq
\label{a1}
\mc{H}_\mr{PF}^{(m)}=\sum_{1\leq i< j \leq N}
\frac{1-  P_{ij}^{(m)}}{(\rho_i-\rho_j)^2}  \, ,
\eeq
where 
$P_{ij}^{(m)}$ denotes the exchange operator 
which interchanges the spins of the $i$-th and $j$-th lattice sites,
and $\rho_i$ is the $i$-th zero point of the 
Hermite polynomial of degree $N$. Similar to the case of 
HS spin model, the PF chain \eq{a1}  
exhibits $Y(gl_m)$ Yangian quantum group symmetry 
for any value of $N$~\cite{Hi95npb}.
Consequently, the  energy eigenvalues
of this spin chain can be expressed through   
the motifs and the corresponding spectrum 
can be reproduced from the energy functions of a one-dimensional classical vertex model. 

Due to the decoupling of  
the spin and coordinate degrees of freedom in the case of
\su{m} spin Calogero model for large values of its coupling constant,  the canonical partition function of the \su{m} PF spin chain \eq{a1} can be derived 
by using the freezing trick.
More precisely, this partition function can be obtained 
by dividing the canonical partition function of the 
\su{m} spin Calogero model through  
that of the spinless Calogero model. 
Thus, for the purpose of deriving the partition function of the \su{m} PF spin chain by using the freezing trick, 
it is necessary to calculate at first the canonical partition function of the \su{m} spin Calogero model. This partition function has been computed in the literature 
by using two different approaches --- a direct one
and an indirect one. 
Polychronakos has originally computed this partition 
function in an indirect way by expanding
the corresponding grand
canonical partition function 
(which can be obtained easily from the  
grand canonical partition function of the spinless Calogero model) as a power series of the fugacity parameter~\cite{Po94}.
Finally, by applying the freezing trick, 
the partition function of the ferromagnetic 
\su{m} PF spin chain \eq{a1} has been derived
in the form
\beq
\label{a2}
\mc{Z}_{A,N}^{(m)}(q)=\sum_{\stackrel{\sum_{i=1}^{m}a_i=N}{a_i\geq 0}}
\qbinom{N}{a_1,a_2,\cdots ,a_m}{q} \, ,
\eeq
where $q\equiv e^{-1/(k_BT)}$, summation is taken over 
all $a_i$ (which are non-negative integers) satisfying the condition $\sum_{i=1}^{m}a_i=N$,
and the $q$-multinomial coefficients are defined as
\[
\qbinom{N}{a_1,a_2,\cdots ,a_m}{q} = 
\frac{(q)_N}{(q)_{a_1}(q)_{a_2}\cdots(q)_{a_m}} \, ,
\]
with $(q)_n\equiv (1-q)(1-q^2)\cdots (1-q^{n})$.
Since each  $q$-multinomial coefficient is a polynomial of $q$,  the partition function $\mc{Z}_{A,N}^{(m)}(q)$ in \eq{a2} 
can also be expressed as a polynomial of $q$.
A supersymmetric generalization of the PF spin chain \eq{a1}, containing both bosonic and fermionic spin degrees of freedom, 
has also been studied and the corresponding partition function has been computed with the help of the freezing trick via  
the indirect approach as described above \cite{BUW99}.
However, it is also possible to directly compute the canonical partition function of the \su{m} spin Calogero model from the 
knowledge of its spectrum. Proceeding in this way and
subsequently applying the freezing trick,  
Barba et al. have derived~\cite{BFGR08epl} the canonical partition function 
of the \su{m} PF spin chain \eq{a1} 
in a form which apparently looks quite different from 
 $\mc{Z}_{A,N}^{(m)}(q)$ in \eq{a2}.

 In this context 
it may be noted that, the classical Rogers-Szeg\"o (RS) polynomial in a single variable
(say, $x$) is defined as $\mbb{H}_N(x,q)=\sum_{k=0}^N 
\qbinom{N}{k,N-k}{q}x^k $ \cite{An76}. 
This RS polynomial has been studied in connection  
with the well known Rogers-Ramanujan identities in number theory. Moreover,   
this RS polynomial can be viewed as a $q$-deformed version 
of the Hermite polynomial, which provides a basis for the coordinate representation of the $q$-oscillator algebra \cite{Ma89,OS08}. 
Different types of homogeneous and inhomogeneous multivariate
generalizations of the classical RS polynomial have also
been studied in the literature. In particular,  
Hikami has observed that the homogeneous 
multivariate RS polynomials (depending on only one type of 
variables) of the form 
\beq
\label{a3}
\mbb{H}_{A,N}^{(m)}(x_1,x_2,\cdots,x_m;q)=
\sum_{\stackrel{\sum_{i=1}^{m}a_i=N}{a_i\geq 0}}
\qbinom{N}{a_1,a_2,\cdots ,a_m}{q} x_1^{a_1} x_2^{a_2} \cdots x_m^{a_m} \, , 
\eeq
reproduce the partition function \eq{a2} of the PF spin chain
in the  limit $x_1=x_2=\cdots =x_m=1$~\cite{Hi95npb,Hi95,Hi97}. Consequently,
a representation of the $Y(gl_m)$ 
Yangian invariant motifs associated 
with the PF spin chain \eq{a1} can be constructed by using  
a recursion relation satisfied by the RS polynomials \eq{a3}.  
It may also be noted that, super RS 
(SRS) polynomials containing two different types of variables  
have been proposed in Ref.~\cite{HB00} for the purpose
of analyzing the spectra and partition functions of the 
supersymmetric PF spin chains 
on the basis of their $Y(gl_{(m|n)})$ super Yangian symmetry.
From the above discussion it is clear that RS
and SRS polynomials 
play an important role in the study of 
PF spin chains and their supersymmetric
generalizations. However, since  
the partition function of the \su{m} PF spin chain obtained 
by using the freezing trick via the direct approach
appears to be quite different from \eq{a2}, such a form of the 
partition function cannot be connected 
with the RS polynomials \eq{a3} in a straightforward way.  

As is well known, root systems associated with Lie algebras 
are widely used in the classification  of 
quantum integrable systems with long-range 
interaction. In particular,  
the above discussed \su{m}  PF spin chains   
with $N$ number of lattice sites
and their supersymmetric generalization are  
related to the 
$A_{N-1}$ root system. However,  
it is  possible to construct exactly solvable 
variants of the PF spin chain \eq{a1} associated with the    
$BC_N$ and $D_N$ 
root systems~\cite{YT96,BFGR08,BFG09}. 
One remarkable feature of the  Hamiltonians 
of the PF spin chains   
associated with the latter root systems is that they contain 
reflection operators like $S_i$ ($i=1,\dots,N$),
which satisfy the relation $S_i^2=\one$ and 
yield a representation of some elements 
appearing in the $BC_N$ or $D_N$ type of Weyl algebra.  
In the special case
when $S_i$ is taken as the spin reversal operator 
$P_i$, which changes the sign of 
the spin component on the $i$-th lattice site,  
the partition functions of  PF spin chains associated with the 
$BC_N$ and $D_N$ root systems have been computed by
using the freezing trick via the direct approach~\cite{BFGR08,BFG09}. Furthermore, by taking 
reflection operators as supersymmetric analogue of   
spin reversal operators (SASRO), partition functions of 
 PF spin chains associated with the  
$BC_N$ root system have been computed by using the 
freezing trick 
via  the direct as well as indirect approaches
\cite{BFGR09}.

However, 
it has been found recently that 
reflection operators can be chosen in 
 more general way
than the above mentioned spin reversal operators and their supersymmetric analogues. 
For example, these reflection operators may be taken 
as arbitrarily polarized spin
reversal operators (PSRO) like $P_i^{(m_1,m_2)}$, 
which acts as the identity on the first $m_1$ elements
of the spin basis and as minus the identity on the rest of the  $m_2$ elements of the spin basis \cite{BBB14}. 
As a result, depending on the action of $P_i^{(m_1,m_2)}$, 
the basis vectors of the $(m_1+m_2)$-dimensional spin space on each 
lattice site can be classified into two groups --- 
$m_1$ elements with 
positive parity and $m_2$ elements with negative parity.
For the special cases like $m_1=m_2$ or  
$m_1=m_2\pm1$, $P_i^{(m_1,m_2)}$ 
reduces to the spin reversal operator
$P_i$ (up to a sign factor) through a similarity transformation.
Choosing the reflection operators as such PSRO,
new exactly solvable spin Calogero models 
of $BC_N$ and $D_N$ type 
have been constructed and the  
partition functions of the related PF chains 
have also been computed 
by using the freezing trick via the direct approach~\cite{BBB14,BDFG15}. 
Furthermore, 
exactly solvable spin Calogero models of $BC_N$ type
have been constructed by taking reflection operators 
as supersymmetric analogues of PSRO (SAPSRO) \cite{BBBD16}.
The strong coupling limit of such spin Calogero models yields  
a large class of 
$BC_N$ type of PF spin chains with SAPSRO, which can   
reproduce all of the previously
studied $BC_N$ type of  PF spin chains at certain limits.

In spite of the above mentioned developments on different variants of the $BC_N$ type of  PF spin chains, 
it is not clear till now whether the spectra of these spin 
chains can be described by some motif like objects 
related to the 
symmetry of these spin chains. Furthermore,  
one may interesting ask whether there exist some one-dimensional classical
vertex models whose energy functions would generate 
the complete spectra of these  $BC_N$ type of  PF spin chains.
However it is known that, 
in the cases of $A_{N-1}$ type of PF spin chains and their 
supersymmetric generalizations, 
multivariate RS and SRS 
polynomials play a key role in solving such problems. 
Hence, as a first step towards solving these
problems for the case of $BC_N$ type of PF spin chains,
in this article our aim is to  
construct the corresponding multivariate SRS polynomials 
and explore some of their properties. 
Since all of the previously
studied  PF spin chains of  $BC_N$ type  
can be obtained by taking 
certain limits of the PF spin chains with SAPSRO, 
it is expected that canonical partition
functions of the later spin chains 
would help us in finding out 
the  general form of the $BC_N$ type of    
multivariate SRS polynomials.  
In this context it may be noted that, the canonical 
partition functions of the $BC_N$ type of PF chains with SAPSRO have been computed earlier by using the freezing trick 
via the direct approach \cite{BBBD16}.
However it may be recalled that, 
for the case of $A_{N-1}$ type of PF spin chains,
 partition functions 
obtained by using the above mentioned procedure  
 cannot be connected 
with the multivariate RS polynomials in a straightforward way.
Therefore, in this article we shall derive a new expression 
for the canonical 
partition functions of the $BC_N$ type of PF chains with SAPSRO
by using the freezing trick via the indirect approach, 
and subsequently use those partition functions 
to construct the corresponding multivariate SRS polynomials.

The arrangement of this paper is as follows. In Sec.~2
we shall briefly review some results of Ref.~\cite{BBBD16}
which are relevant for our purpose,  
like the construction of SAPSRO by using    
the $BC_N$ type of 
Weyl algebra and the method of generating 
PF spin chains with SAPSRO
from the related spin Calogero models 
by applying the freezing trick. Furthermore, we shall    
describe  the Hilbert space 
associated with the $BC_N$ type of spin Calogero models with SAPSRO and the procedure of deriving the spectra of these models 
by choosing a partially ordered set of  basis vectors where 
the corresponding Hamiltonians can be expressed 
in a triangular form. 
In Sec.~3, we shall compute the grand canonical 
partition functions of the $BC_N$ type of ferromagnetic 
as well as anti-ferromagnetic spin
Calogero models with SAPSRO, and expand those   
grand canonical 
partition functions as some power series 
of the fugacity parameter 
to obtain the corresponding canonical 
partition functions. Applying the freezing trick, 
subsequently we shall 
 derive novel expressions for the canonical  
partition functions of the $BC_N$ type of ferromagnetic 
and anti-ferromagnetic PF chains with SAPSRO.
Inspired by the form of such partition functions, 
in Sec.~4 we shall define $BC_N$ type of homogeneous 
multivariate SRS polynomials and also 
find out the corresponding
generating functions. 
Using these generating
functions, we shall  
show that the $BC_N$ type of SRS polynomials
can be expressed as some bilinear 
combinations of the $A_{N-1}$ type of SRS polynomials. 
Furthermore, 
we shall derive a set of recursion relations 
for the partition functions of the $BC_N$ type of PF 
spin chains involving different numbers of lattice sites and 
internal degrees of freedom. Sec.~5 is the concluding section.

\noi \section{$BC_N $ type of spin models with SAPSRO}
\renewcommand{\theequation}{2.{\arabic{equation}}}
\setcounter{equation}{0}
\medskip
It is well known that the 
$BC_N$ type of Weyl algebra is 
generated by the elements like  
$\mc{W}_{ij}$ and $\mc{W}_{i}\, ,$ 
which satisfy the relations 
\bea
&\mc{W}_{ij}^2=\one \, ,
~~~~\mc{W}_{ij}\mc{W}_{jk} =\mc{W}_{ik}\mc{W}_{ij}
=\mc{W}_{jk}\mc{W}_{ik} \, ,~~~~ \mc{W}_{ij}\mc{W}_{kl}
=\mc{W}_{kl}\mc{W}_{ij} \, ,  \nn\\ 
&\mc{W}_{i}^2=\one \, , 
~~~~ \mc{W}_{i}\mc{W}_{j}= \mc{W}_{j}\mc{W}_{i} \, ,
~~~~ \mc{W}_{ij}\mc{W}_{k}=\mc{W}_{k}\mc{W}_{ij} \, , 
~~~~\mc{W}_{ij}\mc{W}_{j}=\mc{W}_{i}\mc{W}_{ij} \, , 
\label{b1}
\eea
where $i,~j,~k,~l\in \{1,2, \cdots, N\}$ are all different indices.  
Representations of this  Weyl algebra play an important  
role in constructing 
$BC_N$ type of quantum integrable 
spin models with long-range interaction. 
For the purpose of describing a class of    
representations of the $BC_N$ type of Weyl 
algebra \eq{b1} on a superspace,  
let us consider a set of operators like
$C_{j \alpha}^\dagger$~($C_{j \alpha}$) which creates (annihilates)
a particle of species $\alpha$ on the $j$-th lattice site.
These creation (annihilation) operators are assumed to be bosonic when $\alpha \in [1,2,....,m]$ and 
fermionic when $\alpha \in [m+1,m+2,....,m+n]$.
Hence, the parity of these operators are defined as 
\bea
 &&\pi(C_{j \alpha})=\pi(C_{j \alpha}^\dagger)=0 ~
\mr{for}~ \alpha \in [1,2,....,m] \, , \nn \\
 &&\pi(C_{j \alpha})=\pi(C_{j \alpha}^\dagger)=1 ~
 \mr{for}~ \alpha \in [m+1,m+2,....,m+n] \, , \nn
\eea
and they satisfy commutation (anti-commutation) relations 
given by 
\beq
[C_{j \alpha},C_{k \beta}]_{\pm}=0 \, ,~ 
[C_{j \alpha}^\dagger,C_{k \beta}^\dagger]_{\pm}=0 \, , ~
[C_{j \alpha},C_{k \beta}^\dagger]_{\pm}=\delta_{jk}\delta_{\alpha \beta} \, ,
\label{b2}
\eeq
where $[C,D]_{\pm} \equiv CD- (-1)^{\pi(C)\pi(D)}DC$.
Let us now consider a finite dimensional
subspace of the related Fock space,  
where each lattice site is occupied by only one particle, i.e., 
$\sum_{\alpha=1}^{m+n} C_{j\alpha}^{\dagger} C_{j\alpha}=1$
for all $j\in \{1,2, \cdots, N\}$.  
The supersymmetric exchange operator 
$\hat{P}_{ij}^{(m|n)}$ is defined on such subspace
of the Fock space as \cite{Ha93} 
\beq
\hat{P}_{ij}^{(m|n)} \equiv
\sum_{\alpha,\beta=1}^{m+n} C_{i \alpha}^\dagger
C_{j \beta}^\dagger C_{i \beta}C_{j \alpha} \, . 
\label{b3}
\eeq

The supersymmetric exchange operator \eq{b3}
can equivalently be
expressed as an operator on the total internal space of $N$ number of spins,   
which is  defined in the following way~\cite{Ba99,BBBD16}. 
Let us denote such total internal space 
as $\mbs{\Sigma}^{(m_1,m_2|n_1,n_2)}$,
where $m_1,~m_2,~n_1,~ n_2$ are some arbitrary non-negative integers
satisfying the relations $m_1+m_2=m$ and $n_1+n_2=n$ . 
The space  $\mbs{\Sigma}^{(m_1,m_2|n_1,n_2)}$ is spanned 
by orthonormal state vectors of the form  $\ket{s_1,\cdots,
s_i, \cdots, s_N}$, 
where 
$s_i \equiv (s_i^1,s_i^2,s_i^3)$ has
three components  
which take discrete values like   
$s_i^1\equiv \pi(s_i) \in \{0,1\} $,  $s_i^2\equiv f(s_i) 
\in \{0,1\} $, and 
\beq 
s_i^3 \in \left \{ \hskip -.47 cm  
\begin{array}{llll} 
&\mbox{~ $\{1,2, \cdots,m_1 \},
~\mr{if}~\pi(s_i)=0~ \mr{and} ~f(s_i)=0$, } \\
&~\mbox{~$\{1, 2, \cdots, m_2 \},
~\mr{if}~\pi(s_i)=0 ~ \mr{and} ~ f(s_i)=1 $, } \\
&~\mbox{~$\{1, 2, \cdots, n_1 \},
~~\mr{if}~\pi(s_i)=1~ \mr{and} ~f(s_i)=0 $, } \\
&~\mbox{~$\{1, 2, \cdots, n_2\},
~~\mr{if}~\pi(s_i)=1~ \mr{and} ~f(s_i)=1 $. } 
\end{array} 
\right.
\label{b4}
\eeq 
Hence, each local spin vector
$s_i$ may be chosen in $(m+n)$ number 
of different ways and  
 $\mbs{\Sigma}^{(m_1,m_2|n_1,n_2)}$
can be expressed in a direct product form given by    
\beq
\mbs{\Sigma}^{(m_1,m_2|n_1,n_2)} 
\equiv \underbrace{\mc{C}_{m+n} 
\otimes \mc{C}_{m+n} 
\otimes \cdots \otimes \mc{C}_{m+n}}_{N} \, ,  
\label{b5}
\eeq
where $\mc{C}_{m+n}$ 
denotes an $(m+n)$-dimensional complex vector space.
It is evident that this $\mbs{\Sigma}^{(m_1,m_2|n_1,n_2)} $
is isomorphic to the subspace of the Fock space,
on which $\hat{P}_{ij}^{(m|n)}$ in \eq{b3} is defined.
A supersymmetric spin exchange 
operator $P_{ij}^{(m|n)}$ is defined  
on the space $\mbs{\Sigma}^{(m_1,m_2|n_1,n_2)}$ as 
\beq
P_{ij}^{(m|n)}\ket{s_1,\cdots,s_i,\cdots,s_j,\cdots,s_N}
=(-1)^{\alpha_{ij}(\mbf{s})}
\ket{s_1,\cdots,s_j,\cdots,s_i,\cdots,s_N},
\label{b6}
\eeq
where
$\alpha_{ij}(\mbf{s})
=\pi(s_i)\pi(s_j)+\left(\pi(s_i)+\pi(s_j)\right)\, 
h_{ij}(\mbf{s})$
and
$h_{ij}(\mbf{s})=\sum_{k=i+1}^{j-1}\pi(s_k)$. 
From \Eq{b6} it follows that, 
the exchange of two spins with $\pi(s_i)=\pi(s_j)=0$
or $\pi(s_i)=\pi(s_j)=1$ produces
a phase factor of $1$ or  $-1$ respectively.
So we may call $s_i$ as a `bosonic' spin  if 
$s_i^1\equiv\pi(s_i)=0$ and a `fermionic' spin if
$s_i^1\equiv\pi(s_i)=1$.
However, it should be noted the exchange one bosonic spin with one fermionic
spin (or, vice versa) produces a nontrivial phase factor of 
$(-1)^{h_{ij}(\mbf{s})}$, where $h_{ij}(\mbf{s})$ represents
the number of fermionic spins within the $i$-th and $j$-th 
lattice sites. 
Using the commutation (anti-commutation) relations in \eq{b2}, 
it can be shown that $\hat{P}_{ij}^{(m|n)}$
in \eq{b3} is completely equivalent to  
$P_{ij}^{(m|n)}$ in \eq{b6}~\cite{Ba99}. 
The action of SAPSRO 
(denoted by $P_i^{(m_1,m_2|n_1,n_2)}$) is defined on the space $\mbs{\Sigma}^{(m_1,m_2|n_1,n_2)}$ as~\cite{BBBD16}
\beq
P_i^{(m_1,m_2|n_1,n_2)}\ket{s_1,\cdots,s_i,\cdots, s_N}
=(-1)^{f(s_i)}\ket{s_1,\cdots,s_i,\cdots, s_N}.
\label{b7}
\eeq
Hence, the 
second component of the spin $s_i$ 
determines its parity  
 under the action of SAPSRO.
It is easy to verify that 
$P_{ij}^{(m|n)}$ in \eq{b6}  
and $P_i^{(m_1,m_2|n_1,n_2)}$ in \eq{b7}  respectively
yield  representations 
of the elements $\mc{W}_{ij}$ and $\mc{W}_{i}\,$   
appearing in the $BC_N$ type of Weyl algebra \eq{b1}.
These representations of the Weyl algebra 
can be used to construct  a large 
class of exactly solvable $BC_N$
type of ferromagnetic PF spin chains with Hamiltonians given by 
\beq
\mathcal{H}^{(m_1,m_2|n_1,n_2)}
=\sum_{i\neq j}\left[\frac{1- P_{ij}^{(m|n)}}{(\xi_i-\xi_j)^2} +
\frac{1- \widetilde{{P}}_{ij}^{(m_1,m_2|n_1,n_2)}}{(\xi_i+\xi_j)^2}\right]
+\beta\sum_{i=1}^{N}\frac{1- P_i^{(m_1,m_2|n_1,n_2)} }{\xi_i^2} \, ,  
\label{b8}
\eeq
where 
 $\beta>0$ is a real parameter,
 $\xi_i=\sqrt{2y_i}$ with $y_i$ 
being the $i$-th zero point  
of the generalized Laguerre polynomial $L_N^{\beta -1}$, and 
$\widetilde{P}_{ij}^{(m_1,m_2|n_1,n_2)} \equiv 
 P_i^{(m_1,m_2|n_1,n_2)}P_j^{(m_1,m_2|n_1,n_2)} P_{ij}^{(m|n)}$.
It may be noted that, the Hamiltonian \eq{b8}
can reproduce all of the previously studied $BC_N$ type of
PF spin chains 
for some specific values of the discrete parameters 
$m_1,~m_2,~n_1$ and $n_2$. For example, 
in the presence of only bosonic or fermionic spins, i.e., when 
  either $n_1=n_2=0$ or $m_1=m_2=0$, 
$\mathcal{H}^{(m_1,m_2|n_1,n_2)}$ 
reduces to the non-supersymmetric PF spin chain 
associated with PSRO~\cite{BBB14}. 
In another special case, where  
the discrete parameters in \eq{b8}
satisfy the relations 
\beq
m_1= \frac{1}{2}\left( m+ \ep \, \tilde{m} \right),~ 
m_2= \frac{1}{2}\left( m - \ep \, \tilde{m} \right),~
n_1= \frac{1}{2}\left( n+ \ep' \, \tilde{n} \right), ~
n_2= \frac{1}{2}\left( n - \ep' \, \tilde{n} \right),
\label{con}
\eeq
with $\ep,\ep^{\prime}=\pm 1$, $\tilde{m}
\equiv m~ \mr{mod}~ 2$ and
$\tilde{n} \equiv n~ \mr{mod}~ 2$,
the exactly solvable Hamiltonian (which depends 
on the parameters $m,n,\ep,\ep'$)
of the $BC_N$ type of 
PF spin chains with SASRO ~\cite{BFGR09} can be obtained from 
$\mathcal{H}^{(m_1,m_2|n_1,n_2)}$ 
through a unitary transformation~\cite{BBBD16}.

Applying the freezing trick,  
 the  Hamiltonians
\eq{b8} of the $BC_N$ type of PF spin chains with SAPSRO
can be derived
from those of $BC_N$ type of spin Calogero models 
containing both coordinate and spin degrees of freedom.  
The Hamiltonians of such spin Calogero models 
are given by 
\bea
H^{(m_1,m_2|n_1,n_2)}=-\sum_{i=1}^{N}\frac{\d^2}{\d x_i^2} 
+\frac{a^2}{4} r^2
+a\sum_{i\neq j} \left[\frac{a-  P_{ij}^{(m|n)}}
{(x_{ij}^-)^2} +
\frac{a-  \widetilde{P}_{ij}^{(m_1,m_2|n_1,n_2)}}{(x_{ij}^+)^2}\right] \nn \\
+\beta a\sum_{i=1}^{N}\frac{\beta a - P_i^{(m_1,m_2|n_1,n_2)}}{x_i^2}\, ,
\label{b9}
\eea
where $a> \frac{1}{2}$ is a real 
coupling constant, $x_{ij}^- \equiv x_i-x_j$, 
$x_{ij}^+ \equiv x_i+x_j$ and $r^2\equiv \sum_{i=1}^N x_i^2$.
The coefficient of the $a^2$ order term in 
the r.h.s. of \eq{b9} may be written as 
\beq
U(\mbf{x})=
\sum_{i\neq j}\left[\frac{1}{(x_{ij}^-)^2}+\frac{1}{(x_{ij}^+)^2}\right]
+\beta^2\sum_{i=1}^{N}\frac{1}{x_i^2}+\frac{r^2}{4}.
\label{b10}
\eeq
Since this $a^2$ order term in $H^{(m_1,m_2|n_1,n_2)}$ dominates 
in the strong coupling limit $a\to \infty $,
the particles of this spin Calogero model 
concentrate at the  minimum $\mbs{\xi}$ of the potential 
$U(\mbf{x})$. Consequently, 
the coordinate and spin degrees of freedom 
of these particles 
decouple from each other. 
Furthermore one can show that, within the configuration space of the Hamiltonian \eq{b9}, the coordinates 
$\xi_i$ of the unique minimum $\mbs{\xi}$ of the potential 
$U(\mbf{x})$ are given by $\xi_i=\sqrt{2y_i}$, where $y_i$'s  
denote the zeros of the generalized Laguerre polynomial 
$L_N^{\beta -1}$ \cite{CS02}. Consequently, 
$H^{(m_1,m_2|n_1,n_2)}$ in \eq{b9} can be written
in $a\to \infty$ limit as 
\beq
H^{(m_1,m_2|n_1,n_2)}
\approx
H_{sc}+a\, \mathcal{H}^{(m_1,m_2|n_1,n_2)}
\, ,
\label{b11}
\eeq
where $H_{sc}$ is the scalar (spinless) Calogero model 
of $BC_N$ type given by 
\beq
H_{sc}=-\sum_{i=1}^N \frac{\partial^2}{\partial x_i^2}
+\frac{a^2}{4}\,r^2
+a(a-1)\sum_{i\neq 
j}\bigg[
\frac{1}{(x_{ij}^-)^2}+\frac{1}{(x_{ij}^+)^2}\bigg]
+\sum_{i=1}^N\frac{\beta a( \beta a -1)}{x_i^2}\, .
\label{b12}
\eeq 
Thus the Hamiltonians \eq{b8} of the $BC_N$
type of PF spin chains with SAPSRO emerge naturally
from the strong coupling limit of the corresponding
spin Calogero models \eq{b9}. Due to \Eq{b11}, 
the eigenvalues of ${H}^{(m_1,m_2|n_1,n_2)}$ 
satisfy the relation
\beq
E_{ij}^{(m_1,m_2|n_1,n_2)} 
\simeq E_i^{sc} + a \, \mc{E}_j^{(m_1,m_2|n_1,n_2)} \, ,
\label{b13}
\eeq
where $E_i^{sc}$ and $\mc{E}_j^{(m_1,m_2|n_1,n_2)}$ 
are two  eigenvalues
of $H_{sc}$ and $\mathcal{H}^{(m_1,m_2|n_1,n_2)}$ respectively. 
With the help of \Eq{b13}, one can derive an exact formula 
for the canonical partition function
$\mc{Z}_{B,N}^{(m_1,m_2|n_1,n_2)}(T)$ of the 
$BC_N$ type of PF spin chain \eq{b8} 
at a given temperature $T$ as 
\beq
\mc{Z}_{B,N}^{(m_1,m_2|n_1,n_2)}(T)=\lim_{a \rightarrow \infty}
 \frac{Z^{(m_1,m_2|n_1,n_2)}_{B,N}(aT)}{Z_{B,N}(aT)} \, ,
\label{b14}
\eeq
where $Z_{B,N}(aT)$ and
$Z_{B,N}^{(m_1,m_2|n_1,n_2)}(aT)$ represent canonical partition functions (at the temperature $aT$)
of the $BC_N$ type of spinless Calogero model (\ref{b12}) and spin Calogero model (\ref{b9})
respectively.

The $BC_N$ type of spinless Calogero model (\ref{b12})
is a well known exactly solvable system  
with ground state wave function of the form    
\beq
\mu({\bf x})=e^{-\frac{a}{4}r^2}~\prod_i|x_i|^{\beta a}
~\prod_{i<j}|x_i^2-x_j^2|^a \, , 
\label{b15}
\eeq
and ground state energy given by 
\beq
\label{b16}
E_0=Na\Big(\beta a +a(N-1)+\frac{1}{2} \Big)\, . 
\eeq
An exact expression for the canonical
partition function of the $BC_N$ type of spinless Calogero model (\ref{b12}) has been derived earlier   
as~\cite{BFGR08} 
\beq
Z_{B,N}(aT)=
\frac{1}{\prod\limits_{j=1}^N (1-q^{2j})} \, ,
\label{b17}
\eeq 
where  $q \equiv e^{-1/(k_BT)}$ and 
the  contribution from the 
ground state energy has been ignored 
without any loss of generality.

The exact spectrum of the $BC_N$ type of  
spin Calogero Hamiltonian \eq{b9}
can be computed by expressing it in a triangular form  
while acting on a partially ordered
set of basis vectors of the corresponding 
Hilbert space~\cite{BBBD16}.
As found in the later reference,
the Hilbert space associated with this spin Calogero Hamiltonian 
is the closure of the linear subspace spanned by the 
wave functions of the form 
\beq
\psi_{\bf{r}}^{\bf{s}} 
\equiv 
\psi_{ r_1 ,\ldots, r_i, \ldots, r_j, \ldots , r_N}
^{s_1, \ldots, s_i, \ldots, s_j, \ldots, s_N}
=\Lambda^{(m_1,m_2|n_1,n_2)}
\left (\phi_{\bf{r}} ({\bf x})|{\bf s}\rangle \right) \, ,
\label{b18}
\eeq
where  $r_i$'s are arbitrary non-negative integers, 
$\phi_{\bf{r}}({\bf x}) \equiv \mu({\bf x})\prod_{i=1}^N x_i^{r_i}$,
$\ket{\mbf{s}} \equiv  \ket{s_1,\cdots,s_N}$
represents an arbitrary basis element of 
the spin space  $\mbs{\Sigma}^{(m_1,m_2|n_1,n_2)}$, and $\Lambda^{(m_1,m_2|n_1,n_2)}$ is a completely
symmetric projector
related to the $BC_N$ type of Weyl algebra. 
It can be 
shown that $\psi_{\bf{r}}^{\bf{s}}$'s in \eq{b18}  
satisfy the symmetry conditions  
\beq
\psi_{ r_1 ,\ldots, r_i, \ldots, r_j, \ldots , r_N}
^{s_1, \ldots, s_i, \ldots, s_j, \ldots, s_N}
=(-1)^{\alpha_{ij}(\mbf{s})} \, 
\psi_{r_1 , \ldots, r_j, \ldots, r_i, \ldots, r_N}^
{s_1, \ldots, s_j, \ldots, s_i, \ldots , s_N} \, , 
\label{b19}
\eeq
and 
\beq
\psi_{r_1, \ldots, r_i, \ldots , r_N}^{s_1, \ldots, s_i, 
\ldots, s_N}=(-1)^{r_i+f(s_i)}~
\psi_{r_1, \ldots, r_i, \ldots , r_N}^{s_1, \ldots, s_i, 
\ldots, s_N}
\, ,
\label{b20}
\eeq
where  $\alpha_{ij}(\mbf{s})$ is defined 
 after \eq{b6} and $1\leq i<j \leq N$. 
Due to these symmetry conditions, 
$\psi_{\bf{r}}^{\bf{s}}$'s corresponding to all 
possible values of $\bf{r}$ and $\bf{s}$ do not form a set of 
linearly independent basis vectors for 
the Hilbert space associated with the spin Calogero Hamiltonian $H^{(m_1,m_2|n_1,n_2)}$. 
However, $\psi_{\bf{r}}^{\bf{s}}$'s  in \eq{b18} 
would lead to a complete set of basis vectors if 
the following three conditions are imposed on the
possible values of $\bf{r}$ and $\bf{s}$.

i) The lower index $\mbf{r}$ in $\psi_{\bf{r}}^{\bf{s}}$
is chosen in an ordered form 
which separately arranges its even and odd components 
into two non-increasing sequences:    
\bea
{\mbf r}~\equiv ~(\mbf{r_e}, \mbf {r_o})&=&(\overbrace{2l_1,
\ldots, 2l_1}^{k_1}, \,  \ldots, \,  \overbrace{2l_s, \ldots,
2l_s}^{k_s}, \nn \\ &~& \overbrace{2p_1+1, \ldots, 2p_1+1}^{g_1}, \, 
\ldots, \, \overbrace{2p_t+1,
\ldots, 2p_t+1}^{g_t}) \, ,
\label{b21}
\eea
where  $0\leq s,\, t \leq N$, 
$l_1>l_2>\ldots>l_s\geqslant0$ and $p_1>p_2>\ldots>p_t\geqslant0$. 
Since any given $\mbf{r}$ can be brought 
 in the ordered form \eq{b21} 
through an appropriate permutation of its 
components, one can choose this ordered form 
as a consequence of the symmetry condition \eq{b19}.
\vskip 2mm
ii) 
The second component of $s_i$ corresponding to each $r_i$ 
is given by   
\beq
s_i^2\equiv f(s_i) =\left \{
\begin{array}{ll} 0 , & \mbox{for} ~r_i\in \mbf {r_e} \,
, \\
1 \, , &
\mbox{for}~ r_i\in \mbf {r_o} \, . 
\end{array} 
\right. 
\label{b22}
\eeq
This is a direct consequence of
the symmetry condition \eq{b20}.
\vskip 2mm
iii) If $r_i=r_j$ for $i<j$, then from \Eq{b22} it follows
that the second components
of the corresponding spins $s_i$ and $s_j$ must have the 
same value. In that case, one can further apply
Eq.~\eq{b19} to obtain an ordering among 
 $s_i$ and $s_j$, 
 by using the rule~$\pi(s_i)\leqslant \pi(s_j)$,  and subsequently, for the case $\pi(s_i)=\pi(s_j)$, by using 
the rule~$s_i^3 \geqslant s_j^3 +\pi(s_j)$.

\vskip 2mm
All $\psi_{\bf{r}}^{\bf{s}}$'s
in \eq{b18}, 
satisfying the above
mentioned three conditions, represent a set of 
(non-orthonormal) basis
vectors of the Hilbert space associated with the spin Calogero Hamiltonian in \eq{b9}.
If a partial ordering is defined among
these basis vectors like 
 $\psi_{\bf{r}}^{\bf{s}}>\psi_{\bf{r}'}^{ \bf{s}'} \, $, 
 for~$|\bf{r}|>|\bf{r}'|$, where 
 $|\mbf{r}| 
 \equiv \sum_{i=1}^N r_i $, 
it can be shown that $H^{(m_1,m_2|n_1,n_2)}$ in \eq{b9} acts  
as an upper triangular matrix on such partially ordered basis vectors: 
\beq
H^{(m_1,m_2|n_1,n_2)}\, \psi_{\bf{r}}^{\bf{s}}=E_{\mbf{r}}^{\mbf{s}} \,
 \psi_{\bf{r}}^{\bf{s}} +\sum_{|\bf{{r}'}|<|\bf{r}|}~C_{\mbf{r'r}} \, 
\psi_{\bf{{r}'}}^{\bf{s'}} \, ,
\label{b23}
\eeq
where $C_{\mbf{r'r}}$'s are real constants,  
$\mbf{s}'$ is a suitable permutation of $\mbf{s}$ and 
\beq
E_{\mbf{r}}^{\mbf{s}} =a|{\bf{r}}|+E_0 \, . 
\label{b24}
\eeq  
Consequently, 
all eigenvalues in the spectrum of $H^{(m_1,m_2|n_1,n_2)}$   
are given by $E_{\mbf{r}}^{\mbf{s}}$ in \eq{b24}, 
where the quantum numbers $\mbf{r}$ and $\mbf{s}$
satisfy the conditions i)-iii). 
Since the r.h.s. of \eq{b24} does not
depend on the quantum number $\mbf{s}$, $E_{\mbf{r}}^{\mbf{s}}$
has an `intrinsic degeneracy' 
which is obtained by counting
the number of all possible choice of 
spin degrees of freedom corresponding to a given $\mbf{r}$.
By using the energy levels \eq{b24} and corresponding
intrinsic degeneracy factors, 
it is possible to directly compute the canonical partition
function $Z_{B,N}^{(m_1,m_2|n_1,n_2)}(aT)$
of the spin Calogero model \eq{b9}.  Furthermore, by 
inserting such expression of 
$Z_{B,N}^{(m_1,m_2|n_1,n_2)}(aT)$ and $Z_{B,N}(aT)$ given in \eq{b17} to the relation \eq{b14}, 
 one can evaluate the canonical partition function 
$\mc{Z}_{B,N}^{(m_1,m_2|n_1,n_2)}(T)$ 
of the spin chains \eq{b8} \cite{BBBD16}.

However the partition 
functions of the $BC_N$ type of PF spin chains with SAPSRO,
obtained in the above mentioned way, have a rather complicated 
form which can not be expressed through the $q$-multinomial coefficients in a straightforward way.
Hence, for the purpose of constructing $BC_N$ type of multivariate SRS polynomials,
in the next section we shall derive a new expression for 
the canonical partition 
functions of the $BC_N$ type of PF spin chains \eq{b8}
through the indirect approach.
More precisely, we shall first compute the grand canonical 
partition functions of the $BC_N$ type of spin
Calogero models with SAPSRO \eq{b9} and expand such  
grand canonical 
partition functions as a power series of the fugacity parameter 
to obtain the corresponding the canonical 
partition functions. Substitution of those canonical partition
functions to the relation \eq{b14} 
 would lead to the desired expressions
 for the canonical partition functions  
of the spin chains \eq{b8}.

\noi \section{
Partition functions of $BC_N$ type of
spin models with SAPSRO}
\renewcommand{\theequation}{3.{\arabic{equation}}}
\setcounter{equation}{0}
\medskip

A remarkable feature of the grand canonical 
partition functions associated with the $A_{N-1}$ type of spin
Calogero models (with harmonic confining potentials) and their 
supersymmetric generalizations 
is that such partition functions 
can be expressed as some simple  
products of the corresponding `basic modes'. 
For example, the grand canonical 
partition function  
 $\mbb{Z}^{(m|0)}_{A}$ of the $m$-flavor bosonic spin
Calogero model 
can be written through the  
corresponding basic mode, i.e., 
the grand canonical
partition function of the one-flavor (spinless) bosonic
Calogero model as~\cite{Po94} 
\beq
\mbb{Z}^{(m|0)}_{A}= \left(\mbb{Z}^{(1|0)}_{A} \right)^m \, . 
\label{c1}
\eeq
It is well known that, up to a constant shift of all 
energy levels, the spectrum of  one-flavor bosonic
Calogero model  of $A_{N-1}$ type coincides with that 
of the $N$ number of free bosonic oscillators. 
Dropping the zero-point energy of these oscillators
and using an identity given by   
\beq
\sum_{k_1\geq k_2 \geq \cdots \geq k_{N}\geq 0} \, 
 q^{\sum\limits_{i=1}^N k_i} =
\frac{1}{(q)_N} \, ,
\label{c2}
\eeq
the canonical partition function ${Z}^{(1|0)}_{A,N}(q)$
of such one-flavor bosonic
Calogero model 
with $N$ number of particles 
can be obtained as   
\beq
{Z}^{(1|0)}_{A,N}(q) = 
\frac{1}{(q)_N} \, .
\label{c3}
\eeq
As a result, the corresponding grand canonical partition function may be expressed as 
\beq 
\mbb{Z}^{(1|0)}_{A}= \sum_{N=0}^{\infty} y^N \cdot 
{Z}^{(1|0)}_{A,N}(q)
= \, \sum_{N=0}^{\infty} \frac{y^N }{(q)_N} \, ,
\label{c4}
\eeq
where $y\equiv q^{-\mu}$ (with $\mu$ being the chemical potential) denotes the fugacity parameter.
Inserting this expression of $\mbb{Z}^{(1|0)}_{A}$
into \Eq{c1}, 
one can derive the grand canonical 
partition function of the $m$-flavor bosonic spin
Calogero model~\cite{Po94}. 

The grand canonical 
partition function of the \su{m|n} supersymmetric spin
Calogero model of $A_{N-1}$ type 
can also be written 
as the product of two types of basic modes as~\cite{BUW99} 
\beq
\mbb{Z}^{(m|n)}_{A}= \left(\mbb{Z}^{(1|0)}_{A} \right)^m 
\left({\mbb{Z}}^{(0|1)}_{A} \right)^n
\, , 
\label{c5}
\eeq
where ${\mbb{Z}}^{(0|1)}_{A}$ represents 
the grand canonical
partition function of the one-flavor (spinless) fermionic
Calogero model. By using the identity 
\beq
\sum_{k_1>k_2 >\cdots > k_N \geq 0}\,  
q^{\sum\limits_{i=1}^N k_i} =
q^{\frac{N(N-1)}{2}} \cdot \frac{1}{(q)_N} \, ,
\label{c6}
\eeq
the canonical partition function 
of such Calogero model with $N$ number of particles 
can be derived as  
\beq
 {Z}^{(0|1)}_{A,N}(q)= 
q^{\frac{N(N-1)}{2}} \cdot \frac{1}{(q)_N} \, ,
\label{c7}
\eeq
and the corresponding grand canonical partition function
 may be obtained as 
\beq 
{\mbb{Z}}^{(0|1)}_{A}
= \sum_{N=0}^{\infty} y^N {Z}^{(0|1)}_{A,N}(q)
= \, \sum_{N=0}^{\infty} y^N \, q^{\frac{N(N-1)}{2}} 
\cdot \frac{1}{(q)_N}. 
\label{c8}
\eeq
 Inserting  $\mbb{Z}^{(1|0)}_{A}$ in \eq{c4}
 and ${\mbb{Z}}^{(0|1)}_{A}$ in \eq{c8} 
into  \Eq{c5}, 
one can derive the grand canonical 
partition function of the \su{m|n} supersymmetric spin
Calogero model of $A_{N-1}$ type~\cite{BUW99}. 

It may be noted that, 
grand canonical partition functions of the $BC_N$ type of 
spin Calogero models with SASRO have been computed 
earlier in Ref.~\cite{BFGR09}.
Those spin Calogero models with SASRO may be considered 
as some special cases of the $BC_N$ type of spin Calogero models with SAPSRO \eq{b9}, since the former models can be obtained from the latter ones by imposing 
the condition \eq{con} and also using a unitary transformation. 
However it has been found in the later reference that, 
instead of only $BC_N$ type
of basic modes, both $BC_N$ and $A_{N-1}$ types of basic modes appear in the expressions of 
grand canonical partition functions of the spin Calogero models with SASRO.  Such a mixture of two different types of 
basic modes in the 
expression of the grand canonical 
partition function is clearly not suitable  
for our present purpose 
of constructing $BC_N$ type of multivariate RS polynomials. 
In the following, our aim is to derive the grand canonical 
partition functions of the $BC_N$ type of spin
Calogero models with SAPSRO \eq{b9}
as simple  products of only $BC_N$ types
of basic modes. 

In the previous section it has been mentioned  
that  $\psi_{\bf{r}}^{\bf{s}}$'s in \eq{b18}, with indices   
$\mbf{r}$ and $\mbf{s}$
satisfying the rules i)-iii), represent a set of 
(non-orthonormal) basis
vectors of the Hilbert space associated with the spin Calogero Hamiltonian  with SAPSRO \eq{b9}. While these rules 
for ordering $\mbf{r}$ and $\mbf{s}$ 
are very convenient for  
computing the canonical partition function 
of the Hamiltonian \eq{b9}, 
they are not suitable for computing the corresponding 
grand canonical partition function and 
they do not uniquely follow
from the symmetry conditions \eq{b19} and \eq{b20}. 
Indeed, for the purpose of  computing the grand canonical partition function of the Hamiltonian \eq{b9} from its spectrum,  
it is necessary to order at first the upper index $\mbf{s}$ of 
$\psi_{\bf{r}}^{\bf{s}}$ in an appropriate way 
and subsequently find out the rules which the lower index
$\mbf{r}$ should
obey. Hence, instead of using the rules i)-iii), 
we order the  indices $\mbf{s}$  and $\mbf{r}$ of the state vectors  \eq{b18}
by using the following equivalent set of rules to obtain 
essentially the same set of complete basis vectors:
  \vskip .2 cm 
1)~Let us define the difference between two 
local spin vectors  $s \equiv (s^1,s^2,s^3)$
and $\bar{s} \equiv (\bar{s}^1,\bar{s}^2,\bar{s}^3)$
as ${s}-\bar{s}=(s^1-\bar{s}^1,s^2-\bar{s}^2, s^3-\bar{s}^3 )$, 
and assume that $s<\bar{s}$ if the first nonvanishing
component of ${s}-\bar{s}$ is negative. 
Using the symmetry condition \eq{b19}, we order the index 
 $\mbf{s} \equiv (s_1, s_2, \cdots ,s_N)$ such that 
 $s_i\leq s_j$ for $i<j$. 
\vskip .11 cm  
2)~If $s_i=s_j$ for $i<j$, then by using  \eq{b19}
we order the  components $r_i$ and $r_j$ within   
$\mbf{r} \equiv (r_1, r_2, \cdots ,r_N)$  
such that $r_i\geqslant r_j+\pi(s_i)$. 
\vskip .11 cm
3)~Due to the condition \eq{b20}, $r_i$ is taken as an even
non-negative integer if $s_i^2 \equiv f(s_i)=0$, and $r_i$ 
is taken as an odd positive integer if $s_i^2 \equiv f(s_i)=1$.
\vskip .11 cm
As before, a partial ordering may be defined among
these relabeled basis vectors as:
 $\psi_{\bf{r}}^{\bf{s}}>\psi_{\bf{r}'}^{ \bf{s}'} \, $, 
 for~$|\bf{r}|>|\bf{r}'|$.  
It is evident that, in analogy with \eq{b23}, 
the spin Calogero Hamiltonian  \eq{b9} would act   
as an upper triangular matrix on such partially ordered basis vectors. As a result, 
all eigenvalues in the spectrum of $H^{(m_1,m_2|n_1,n_2)}$   
can equivalently be given by 
$E_{\mbf{r}}^{\mbf{s}}$ in \eq{b24}, 
where the indices $\mbf{r}$ and $\mbf{s}$  
are ordered by using the new set of rules 1)-3). 

Next, we assume that the local
spin $s \equiv (s^1,s^2,s^3)$ occurs 
$\gamma^{s^1,s^2,s^3}$ times in the configuration 
$\mbf{s} \equiv (s_1, s_2, \cdots ,s_N)$. It is evident 
that $N$ can be written as 
\beq 
N=\sum_{s^1,s^2,s^3}
\gamma^{{s^1\!,s^2\!,s^3}} \, .
\label{c9}
\eeq
Using the condition 1), we explicitly order the 
configuration $\mbf{s}$ as
\beq
\mbf{s}=(S_1,S_2,S_3,S_4) \, , 
\label{c10}
\eeq
where 
\bea
&&S_1=\underbrace{(001),\cdots,(001)}_{\gamma^{{001}}},
\cdots\cdots,
\underbrace{(00m_1),\cdots,(00m_1)}_{\gamma^{{00m_1}}} \, ,
\nn \\
&&S_2=\underbrace{(011),\cdots,(011)}_{\gamma^{{011}}},
\cdots\cdots,
\underbrace{(01m_2),\cdots,(01m_2)}_{\gamma^{{01m_2}}} \, ,
\nn \\
&&S_3=\underbrace{(101),\cdots,(101)}_{\gamma^{{101}}},
\cdots\cdots,
\underbrace{(10n_1),\cdots,(10n_1)}_{\gamma^{{10n_1}}} \, ,
\nn \\
&&S_4=\underbrace{(111),\cdots,(111)}_{\gamma^{{111}}},
\cdots\cdots,
\underbrace{(11n_2),\cdots,(11n_2)}_{\gamma^{{11n_2}}} \, . 
\label{c11}
\eea
Let $r_i^{s^1,s^2,s^3}$ be the local lower index corresponding
to the $i$-th copy of the local upper index $s \equiv (s^1,s^2,s^3)$, where $i\in \{1,2, \cdots , \gamma^{s^1,s^2,s^3} \}$.
Due to the condition 2), 
we obtain a restriction on $r_i^{s^1,s^2,s^3}$ as 
\beq 
r_{i}^{s^1,s^2,s^3}\geqslant r_{i+1}^{s^1,s^2,s^3}+s^1\, ,  
\label{c12}
\eeq
where  $i\in \{1,2, \cdots , \gamma^{{s^1,s^2,s^3}}-1 \}$.
Furthermore, by using the condition 3), 
we can express any  $r_i^{s^1,s^2,s^3}$ in the form 
\beq 
r_i^{s^1,s^2,s^3} = 2k_i^{s^1,s^2,s^3} +\delta_{s^2,1} ~ ,
\label{c13}
\eeq
where $k_i^{s^1,s^2,s^3}$ is a non-negative integer.  
Hence, the 
lower index $\mbf{r}$ corresponding to the
upper index $\mbf{s}$ in \eq{c10} and \eq{c11} may be written
as  
\beq
\mbf{r} =(R_1,R_2,R_3,R_4) \, , 
\label{c14}
\eeq
where 
\bea
&&R_1=\underbrace{2{k}^{001}_1, \cdots, 2{k}^{001}_{\gamma^{{001}}}}_{\gamma^{{001}}},
\cdots\cdots,
\underbrace{2{k}^{00m_1}_1, \cdots, 2{k}^{00m_1}_{\gamma^{{00m_1}}}}_{\gamma^{{00m_1}}} \, ,
\nn \\
&&R_2=\underbrace{2{k}^{011}_1+1, \cdots, 2{k}^{011}_{\gamma^{{011}}}+1}_{\gamma^{{011}}},
\cdots\cdots,
\underbrace{2{k}^{01m_2}_1 +1, \cdots, 2{k}^{01m_2}_{\gamma^{{01m_2}}}+1}_{\gamma^{{01m_2}}} \, ,
\nn \\
&&R_3=\underbrace{2{k}^{101}_1, \cdots, 2{k}^{101}_{\gamma^{{101}}}}_{\gamma^{{101}}},
\cdots\cdots,
\underbrace{2{k}^{10n_1}_1, \cdots, 2{k}^{10n_1}_{\gamma^{{10n_1}}}}_{\gamma^{{10n_1}}} \, ,
\nn \\
&&R_4=\underbrace{2{k}^{111}_1+1, \cdots, 2{k}^{111}_{\gamma^{{111}}}+1}_{\gamma^{{111}}},
\cdots\cdots,
\underbrace{2{k}^{11n_2}_1 +1, \cdots, 2{k}^{11n_2}_{\gamma^{{11n_2}}}+1}_{\gamma^{{11n_2}}} 
\, . 
\label{c15}
\eea
Due to Eqs.~\eq{c12} and \eq{c13}, 
we obtain a restriction on $k_i^{s^1,s^2,s^3}$ 
appearing in \Eq{c15} as 
\beq 
k_i^{s^1,s^2,s^3}\geqslant k_{i+1}^{s^1,s^2,s^3}+s^1\, ,
\label{c16}
\eeq
where  $i\in \{1,2, \cdots , \gamma^{{s^1,s^2,s^3}}-1 \}$.

Let us now try to evaluate the grand  canonical partition function of the $BC_N$ type of spin CS model with SAPSRO
\eq{b9}. To this end, we note that
$|\mbf{r} |$ can be written as 
\beq
|\mbf{r}| = 
\sum_{s^1,s^2,s^3}
\sum_{i=1}^{\gamma^{{s^1\!,s^2\!,s^3}}}~
r_i^{s^1\!,s^2\!,s^3} \, . 
\label{c17}
\eeq
By using the above  form of  
$|\mbf{r}|$, the expression of $N$ given in \eq{c9}
and the energy eigenvalue relation \eq{b24},
we define 
the grand  canonical partition function associated with the 
Hamiltonian \eq{b9} as  
\beq 
{\mbb{Z}}^{(m_1,m_2|n_1,n_2)}_{B}(aT, \mu) 
= \sum_{ \{\gamma^{{t^1,t^2,t^3}} \}\, \geq \, 0} ~ 
\acute{\sum\limits_{\{r_j^{t^1,t^2,t^3}\} \, \geq \, 0}}
q^{|\mbf{r}|-\mu N} \ , 
\label{c18}
\eeq
where the symbol
$\sum_{ \{\gamma^{{t^1,t^2,t^3}} \}\, \geq \, 0}$
implies multiple sums over all $\gamma^{{t^1,t^2,t^3}}$ ranging 
from $0$ to $\infty$,
the symbol $\acute{\sum\limits}_{\{r_j^{t^1,t^2,t^3}\} \, \geq \, 0}$ implies 
restricted multiple sums over all $r_j^{t^1,t^2,t^3} $ 
(ranging 
from $0$ to $\infty$) which satisfy the conditions \eq{c12} 
and \eq{c13}, 
and the contribution from the ground state energy has been 
ignored as before.
Using \eq{c13}, one can rewrite $|\mbf{r} |$ in \eq{c17} as 
\[
|\mbf{r}| = \sum_{s^1,s^2,s^3}
\Big\{\sum_{i=1}^{\gamma^{{s^1\!,s^2\!,s^3}}} 
2k_i^{s^1\!,s^2\!,s^3} +\delta_{s_2,1}~ 
\gamma^{{s^1\!,s^2\!,s^3}} \Big\} \, . 
\]
Inserting the above expression of $|\mbf{r} |$ along with 
$N$ given in \eq{c9} into \Eq{c18}, we obtain 
\bea
&&{\mbb{Z}}^{(m_1,m_2|n_1,n_2)}_{B} \nn \\
&&~~~~~~= \sum_{ \{\gamma^{{t^1,t^2,t^3}} \}\, \geq \, 0} ~ 
\acute{\sum\limits_{\{k_j^{t^1,t^2,t^3}\} \, \geq \, 0}}
q^{\, \sum\limits_{s^1\!,s^2\!,s^3}
\left\{\sum\limits_{i=1}^{\gamma^{{s^1\!,s^2\!,s^3}}}
2k_i^{s^1\!,s^2\!,s^3} 
+(\delta_{s_2,1}-\mu)~ \gamma^{{s^1\!,s^2\!,s^3}} 
\right\} } \nn \\
&&~~~~~~= \sum_{ \{\gamma^{{t^1,t^2,t^3}} \}\, \geq \, 0} ~ 
\acute{\sum\limits_{\{k_j^{t^1,t^2,t^3}\} \, \geq \, 0}}~
\prod_{s^1,s^2,s^3}
q^{ \sum\limits_{i=1}^{\gamma^{{s^1\!,s^2\!,s^3}}}
2k_i^{s^1\!,s^2\!,s^3} 
+(\delta_{s_2,1}-\mu)~ \gamma^{{s^1\!,s^2\!,s^3}} } , 
\label{c19}
\eea
 where ${\mbb{Z}}^{(m_1,m_2|n_1,n_2)}_{B} \equiv 
 {\mbb{Z}}^{(m_1,m_2|n_1,n_2)}_{B}(aT, \mu)$, 
 and the symbol 
 $\acute{\sum\limits}_{\{k_j^{t^1,t^2,t^3}\} \, 
 \geq \, 0}$ implies 
restricted multiple sums over all $k_j^{t^1,t^2,t^3} $ 
(ranging 
from $0$ to $\infty$) which satisfy the condition \eq{c16}. 
It is possible to express 
${\mbb{Z}}^{(m_1,m_2|n_1,n_2)}_{B}$ in \eq{c19} 
in  a factorized form like 
\bea
&&{\mbb{Z}}^{(m_1,m_2|n_1,n_2)}_{B}\nn \\
&&~~~~= \prod_{s^1,s^2,s^3} \, 
\sum_{\gamma^{s^1\!,s^2\!,s^3} =0}^{\infty}   
q^{ (\delta_{s_2,1}-\mu)~ \gamma^{{s^1\!,s^2\!,s^3}} }
\left\{  
\acute{\sum\limits_{k_1^{s^1,s^2,s^3},\cdots, \, 
k_{\gamma^{s^1\!,s^2\!,s^3}}^{s^1,s^2,s^3}   \, \geq \, 0}}~
q^{\sum\limits_{i=1}^{\gamma^{{s^1\!,s^2\!,s^3}}}
2k_i^{s^1\!,s^2\!,s^3}  } \right\},~~~~ 
\label{c20}
\eea
where $\acute{\sum}_{k_1^{s^1,s^2,s^3},\cdots, \, 
k_{\gamma^{s^1\!,s^2\!,s^3}}^{s^1,s^2,s^3}   \, \geq \, 0}
$
implies 
restricted multiple sums over the variables
$k_1^{s^1,s^2,s^3}$, $k_2^{s^1,s^2,s^3}$, $ \cdots ,$ $k_{\gamma^{s^1,s^2,s^3}}^{s^1,s^2,s^3}$ 
(ranging 
from $0$ to $\infty$) which satisfy the condition \eq{c16}. 
Let us now rewrite \Eq{c20} in the form  
\beq
{\mbb{Z}}^{(m_1,m_2|n_1,n_2)}_{B} 
= \prod_{s^1,s^2} \left ( \prod_{s_3} 
{\mbb{Z}}^{s^1,s^2,s^3}_{B}\right) \, ,
\label{c21}
\eeq
where ${\mbb{Z}}^{s^1,s^2,s^3}_{B}$ is given by
\beq 
{\mbb{Z}}^{s^1,s^2,s^3}_{B}=
\sum_{\gamma^{s^1\!,s^2\!,s^3} =0}^{\infty} \,  
q^{ (\delta_{s_2,1}-\mu)\, 
\gamma^{s^1\!,s^2\!,s^3} } \left\{
\acute{\sum\limits_{k_1^{s^1,s^2,s^3},\cdots, \, 
k_{\gamma^{s^1\!,s^2\!,s^3}}^{s^1,s^2,s^3}   \, \geq \, 0}}~
q^{\sum\limits_{i=1}^{\gamma^{s^1\!,s^2\!,s^3}}
2k_i^{s^1\!,s^2\!,s^3}  } \right\} \, .  
\label{c22}
\eeq
Even though the above expression of 
${\mbb{Z}}^{s^1,s^2,s^3}_{B}$
implicitly depends on $s_1$ 
through the condition \eq{c16} and 
explicitly depends on  $s_2$, it does not depend 
at all on the value of $s_3$. Consequently, 
by replacing each value of $s_3$ with $1$,
we can express \Eq{c21} in a factorized form like  
\beq
{\mbb{Z}}^{(m_1,m_2|n_1,n_2)}_{B} 
=\left\{ {\mbb{Z}}^{0,0,1}_{B}\right\}^{m_1}
\left\{ {\mbb{Z}}^{0,1,1}_{B}\right\}^{m_2}
\left\{ {\mbb{Z}}^{1,0,1}_{B}\right\}^{n_1}
\left\{ {\mbb{Z}}^{1,1,1}_{B}\right\}^{n_2}
\, . 
\label{c23}
\eeq
For small values of the discrete parameters
satisfying the condition $m_1+m_2+n_1+n_2=1$
(like $m_1=1, m_2=n_1=n_2=0$),   
\Eq{c23} leads to the relations     
\beq
~~~~{\mbb{Z}}^{(1,0|0,0)}_{B}= {\mbb{Z}}^{0,0,1}_{B}, ~
{\mbb{Z}}^{(0,1|0,0)}_{B}= {\mbb{Z}}^{0,1,1}_{B}, ~ {\mbb{Z}}^{(0,0|1,0)}_{B}= {\mbb{Z}}^{1,0,1}_{B}, ~
{\mbb{Z}}^{(0,0|0,1)}_{B}= {\mbb{Z}}^{1,1,1}_{B} .
 ~~\tag 
\label{3.24a,b,c,d}
\eeq
\addtocounter{equation}{1}
Hence, we can write the grand  canonical partition function
\eq{c23}
of the $BC_N$ type of spin CS model \eq{b9}
completely in terms of the corresponding basic modes as
\beq
{\mbb{Z}}^{(m_1,m_2|n_1,n_2)}_{B}
=\left\{ {\mbb{Z}}^{(1,0|0,0)}_{B}  \right\}^{m_1}
\left\{ {\mbb{Z}}^{(0,1|0,0)}_{B} \right\}^{m_2}
\left\{ {\mbb{Z}}^{(0,0|1,0)}_{B}\right\}^{n_1}
\left\{ {\mbb{Z}}^{(0,0|0,1)}_{B}\right\}^{n_2}
\, . 
\label{c25}
\eeq

Let us now try to evaluate the four 
$BC_N$ type of basic modes appearing 
in the above relation. 
Using Eqs.~(3.24a) and \eq{c22}, along with the condition 
\eq{c16} (for the case $s^1=0,s^2=0,s^3=1$), we obtain 
\[
{\mbb{Z}}^{(1,0|0,0)}_{B}=
\sum_{N =0}^{\infty} \, y^N  
\sum\limits_{k_1^{001} \geq k_2^{001} \geq 
\cdots \, \geq k_{N}^{001}   \, \geq \, 0}~
q^{\sum\limits_{i=1}^{N}
2k_i^{001}  } \, .  
\]
Using the identity \eq{c2},  
the above equation can be written as   
\beq 
{\mbb{Z}}^{(1,0|0,0)}_{B}=
\sum_{N =0}^{\infty} \, y^N  \frac{1}{(q^2)_N}
 \, .  
\label{c26}
\eeq
Next, by using Eqs.~(3.24b) and \eq{c22}, along with 
the condition \eq{c16} (for the case $s^1=0,s^2=1,s^3=1$), we obtain
\[
{\mbb{Z}}^{(0,1|0,0)}_{B}=
\sum_{N =0}^{\infty} \, y^N q^N   
\sum\limits_{k_1^{011} \geq k_2^{011} \geq 
\cdots \, \geq k_{N}^{011}   \, \geq \, 0}~
q^{\sum\limits_{i=1}^{N}
2k_i^{011}  } \, .  
\]
Again, using the identity \eq{c2},  
the above equation can be written as 
\beq 
{\mbb{Z}}^{(0,1|0,0)}_{B}=
\sum_{N =0}^{\infty} \, y^N  \frac{q^N}{(q^2)_N}
 \, .  
\label{c27}
\eeq
Next, by using Eqs.~(3.24c) and \eq{c22}, along with 
the condition \eq{c16} (for the case $s^1=1,s^2=0,s^3=1$), we obtain
\[
{\mbb{Z}}^{(0,0|1,0)}_{B}=
\sum_{N =0}^{\infty} \, y^N  
\sum\limits_{k_1^{101} > k_2^{101} > 
\cdots \, > k_{N}^{101}   \, \geq \, 0}~
q^{\sum\limits_{i=1}^{N}
2k_i^{101}  } \, ,  
\]
which, due to the identity \eq{c6},  leads to  
\beq 
{\mbb{Z}}^{(0,0|1,0)}_{B}=
\sum_{N =0}^{\infty} \, y^N\, \frac{q^{N(N-1)}}{(q^2)_N}
 \, .  
\label{c28}
\eeq
Finally, by using Eqs.~(3.24d) and \eq{c22}, along with 
the condition \eq{c16} (for the case $s^1=1,s^2=1,s^3=1$), we obtain
\[
{\mbb{Z}}^{(0,0|0,1)}_{B}=
\sum_{N =0}^{\infty} \, y^N  q^N
\sum\limits_{k_1^{111} > k_2^{111} > 
\cdots \, > k_{N}^{111}   \, \geq \, 0}~
q^{\sum\limits_{i=1}^{N}
2k_i^{111}  } \, ,  
\]
which, due to the identity \eq{c6},  leads to
\beq 
{\mbb{Z}}^{(0,0|0,1)}_{B}=
\sum_{N =0}^{\infty} \, y^N  \frac{q^{N^2}}{(q^2)_N}
 \, .  
\label{c29}
\eeq

The grand  canonical partition function
${\mbb{Z}}^{(m_1,m_2|n_1,n_2)}_{B}$
can be formally expanded as a power series 
of the fugacity parameter $y$ as 
\beq
{\mbb{Z}}^{(m_1,m_2|n_1,n_2)}_{B}
=\sum_{N =0}^{\infty} \, 
y^N \, {Z}^{(m_1,m_2|n_1,n_2)}_{B,N}(aT) \, .  
\label{c30}
\eeq
Inserting 
the expressions of the basic modes 
given in Eqs.~\eq{c26}, \eq{c27}, \eq{c28} and \eq{c29} 
into the r.h.s. of \Eq{c25}, and comparing this r.h.s.
of the latter 
equation with that of \Eq{c30},
we obtain a new expression for 
the canonical partition function of  
the $BC_N$ type of spin CS model \eq{b9} as 
\beq
{Z}^{(m_1,m_2|n_1,n_2)}_{B,N}(aT)
=\sum_{\stackrel{\sum\limits_{i=1}^{m_1} a_i+\sum\limits_{j=1}^{m_2} b_j
+\sum\limits_{k=1}^{n_1} c_k+\sum\limits_{l=1}^{n_2} d_l=N}
{a_i\geq 0,~ b_j\geq 0, ~c_k\geq 0, ~d_l\geq 0}}~
\frac{q^{\sum\limits_{j=1}^{m_2} b_j+\sum\limits_{k=1}^{n_1} c_k(c_k-1)+\sum\limits_{l=1}^{n_2} d_l^2}}
{\prod\limits_{i=1}^{m_1}(q^2)_{a_i}\prod\limits_{j=1}^{m_2}(q^2)_{b_j}\prod\limits_{k=1}^{n_1}(q^2)_{c_k}
\prod\limits_{l=1}^{n_2}(q^2)_{d_l}} \, .
\label{c31}
\eeq 
Inserting the above expression of 
${Z}^{(m_1,m_2|n_1,n_2)}_{B,N}(aT)$, along with 
$Z_{B,N}(aT)$ given in \eq{b17}, into the relation \eq{b14}, 
we also get a new 
expression for the 
canonical partition function of the $BC_N$ type of 
ferromagnetic PF chain  \eq{b8} as
\beq
\mc{Z}^{(m_1,m_2|n_1,n_2)}_{B,N}(q)
=\sum_{\stackrel{\sum\limits_{i=1}^{m_1} a_i+\sum\limits_{j=1}^{m_2} b_j
+\sum\limits_{k=1}^{n_1} c_k+\sum\limits_{l=1}^{n_2} d_l=N}
{a_i\geq 0,~ b_j\geq 0, ~c_k\geq 0, ~d_l\geq 0}}~
\frac{\left(q^2\right)_N \cdot q^{\sum\limits_{j=1}^{m_2} b_j+\sum\limits_{k=1}^{n_1} c_k(c_k-1)+\sum\limits_{l=1}^{n_2} d_l^2}}
{\prod\limits_{i=1}^{m_1}(q^2)_{a_i}\prod\limits_{j=1}^{m_2}(q^2)_{b_j}\prod\limits_{k=1}^{n_1}(q^2)_{c_k}
\prod\limits_{l=1}^{n_2}(q^2)_{d_l}} \, ,
\label{c32}
\eeq 
where, for the sake of convenience,  
the variable $q$ is used  
(instead of $T$) as the argument 
of $\mc{Z}^{(m_1,m_2|n_1,n_2)}_{B,N}$.  It may be noted that, 
$\mc{Z}^{(m_1,m_2|n_1,n_2)}_{B,N}(q)$ in 
\eq{c32} can be rewritten 
by using the $q^2$-multinomial coefficients as
\bea
&&\mc{Z}^{(m_1,m_2|n_1,n_2)}_{B,N}(q) \nn \\
&&=\sum_{\stackrel{\sum\limits_{i=1}^{m_1} a_i+\sum\limits_{j=1}^{m_2} b_j
+\sum\limits_{k=1}^{n_1} c_k+\sum\limits_{l=1}^{n_2} d_l=N}
{a_i\geq 0,~ b_j\geq 0, ~c_k\geq 0, ~d_l\geq 0}}~
 q^{\sum\limits_{j=1}^{m_2} b_j+\sum\limits_{k=1}^{n_1} c_k(c_k-1)+\sum\limits_{l=1}^{n_2} d_l^2}
 \qbinom{N}{\{a\}_{m_1}\{b\}_{m_2}\{c\}_{n_1}\{d\}_{n_2}}{q^2}
 \, ,~~~~~~
\label{c33}
\eea 
where the notations 
$\{a\}_{m_1}\equiv a_1,\cdots ,a_{m_1}$,
$\{b\}_{m_2}\equiv b_1,\cdots ,b_{m_2}$,
$\{c\}_{n_1} \equiv c_1,\cdots ,c_{n_1}$ and
$\{d\}_{n_2} \equiv d_1,\cdots ,d_{n_2}$ are used. 

We like to make a comment here on the rather surprising
appearance of both $BC_N$ and $A_{N-1}$ types of basic modes 
in the grand canonical partition functions of the 
$BC_N$ type of 
spin Calogero models with SASRO, as found in Ref.~\cite{BFGR09}.  
Choosing $m$ as any odd number, $n$  as any even number
and taking two discrete parameters as $\ep=\ep'=1$, 
the grand canonical partition function 
${\mbb{Z}}^{(m|n)}_{11}$
of the $BC_N$ type of 
spin Calogero model with SASRO has been computed 
in the latter reference as 
\beq
{\mbb{Z}}^{(m|n)}_{11}
={\mbb{Z}}^{(1,0|0,0)}_{B} \,   
\left\{ {\mbb{Z}}^{(1|0)}_{A} \right\}^{\frac{m-1}{2}}
\left\{ {\mbb{Z}}^{(0|1)}_{A}\right\}^{\frac{n}{2}}
\, , 
\label{c34}
\eeq
where we have 
used the notations of the present paper
for all basic modes appearing in the 
r.h.s. of the above equation.
Since spin Calogero models with SASRO can be reproduced from the more general class of spin Calogero models with SAPSRO \eq{b9}
by imposing the condition \eq{con}, it should 
be possible to obtain 
 the grand canonical partition functions of the former models 
 from those of the later models by imposing the same 
 condition. If $m$ is  an odd number, $n$ is an even number
 and $\ep=\ep'=1$, 
the condition \eq{con} yields $m_1=(m+1)/2$, $m_2=(m-1)/2$ and
$n_1=n_2=n/2$. Substituting these values of 
$m_1,~m_2,~n_1$ and $n_2$ in  \Eq{c25}, 
and also replacing the corresponding $ {\mbb{Z}}^{(m_1,m_2|n_1,n_2)}_{B}$ with the notation ${\mbb{Z}}^{(m|n)}_{11}$,
we find that 
\beq
{\mbb{Z}}^{(m|n)}_{11}
= {\mbb{Z}}^{(1,0|0,0)}_{B}  
\left\{ {\mbb{Z}}^{(1,0|0,0)}_{B}{\mbb{Z}}^{(0,1|0,0)}_{B} \right\}^{\frac{m-1}{2}}
\left\{ {\mbb{Z}}^{(0,0|1,0)}_{B} {\mbb{Z}}^{(0,0|0,1)}_{B}\right\}^{\frac{n}{2}}
\, . 
\label{c35}
\eeq
Equating the r.h.s. of \Eq{c34} with that of \Eq{c35}, we 
obtain novel relations like  
\beq
{\mbb{Z}}^{(1|0)}_{A}= 
{\mbb{Z}}^{(1,0|0,0)}_{B}{\mbb{Z}}^{(0,1|0,0)}_{B} \, ,~~~~ 
{\mbb{Z}}^{(0|1)}_{A}= 
{\mbb{Z}}^{(0,0|1,0)}_{B}{\mbb{Z}}^{(0,0|0,1)}_{B} \, , 
\tag
\label{3.36a,b}
\eeq
\addtocounter{equation}{1}
which connects the $A_{N-1}$ and $BC_N$
types of basic modes associated with the  
grand canonical partition functions of the corresponding 
 one-flavor Calogero models. 
 Inserting the expressions of ${\mbb{Z}}^{(1|0)}_{A}$ in 
 \eq{c4}, ${\mbb{Z}}^{(1,0|0,0)}_{B}$ in \eq{c26}, and  
 ${\mbb{Z}}^{(0,1|0,0)}_{B}$ in \eq{c27} into  
 Eq.~(3.36a), and comparing the coefficients of $q^N$ from both 
 sides of the later equation, we obtain a $q$-identity of the form  
 \beq
 \frac{1}{(q)_N}=\sum_{r=0}^N \frac{q^{N-r}}{(q^2)_{r}\cdot 
 (q^2)_{N-r}} \, .
 \label{c37}
 \eeq
 Similarly, inserting the expressions of 
 ${\mbb{Z}}^{(0|1)}_{A}$ in 
 \eq{c8}, ${\mbb{Z}}^{(0,0|1,0)}_{B}$ in \eq{c28}, and  
 ${\mbb{Z}}^{(0,0|0,1)}_{B}$ in \eq{c29} into 
 Eq.~(3.36b), and comparing the coefficients of $q^N$ from both 
 sides of the later equation, we obtain another $q$-identity of the form  
 \beq
 \frac{1}{(q)_N}=\sum_{r=0}^N 
\frac{q^{\frac{1}{2}(N-2r)(N-2r+1)}}
{(q^2)_{r} \cdot (q^2)_{N-r}} \, .
\label{c38}
 \eeq
 One can easily verify that,
 for all possible choice of the parameters $m,n, \ep $ 
 and $ \ep'$, the grand canonical
 partition functions  of 
 the spin Calogero models with SASRO~\cite{BFGR09}
can be reproduced in a similar way from \Eq{c25}  
by using the condition \eq{con} and the 
relations (3.36a,b). Hence, the 
appearance of both $BC_N$ and $A_{N-1}$ types of basic modes 
in the grand canonical partition functions of the 
$BC_N$ type of 
spin Calogero models with SASRO can be explained 
by employing the relations (3.36a,b).

It may be noted that,
following a procedure similar to the case of $BC_N$ type of 
ferromagnetic PF chain \eq{b8}, one can also calculate the 
partition function of the $BC_N$ type of 
anti-ferromagnetic PF chain with Hamiltonian given by 
\beq
\wt{\mc{H}}^{(m_1,m_2|n_1,n_2)}
=\sum_{i\neq j}\left[\frac{1+ P_{ij}^{(m|n)}}{(\xi_i-\xi_j)^2} +
\frac{1+ \widetilde{{P}}_{ij}^{(m_1,m_2|n_1,n_2)}}{(\xi_i+\xi_j)^2}\right]
+\beta\sum_{i=1}^{N}\frac{1+ P_i^{(m_1,m_2|n_1,n_2)} }{\xi_i^2} \, .   
\label{c39}
\eeq
By using the freezing trick, the above Hamiltonian   
can be obtained from the $BC_N$ type of spin Calogero Hamiltonian like  
 \bea
\wt{H}^{(m_1,m_2|n_1,n_2)}=-\sum_{i=1}^{N}\frac{\d^2}{\d x_i^2} 
+\frac{a^2}{4} r^2
+a\sum_{i\neq j} \left[\frac{a + P_{ij}^{(m|n)}}
{(x_{ij}^-)^2} +
\frac{a +  \widetilde{P}_{ij}^{(m_1,m_2|n_1,n_2)}}{(x_{ij}^+)^2}\right] \nn \\
+\beta a\sum_{i=1}^{N}\frac{\beta a + P_i^{(m_1,m_2|n_1,n_2)}}{x_i^2}\, .
\label{c40}
\eea
Hence, we can derive an analogue of \Eq{b14} as 
\beq
\wt{\mc{Z}}_{B,N}^{(m_1,m_2|n_1,n_2)}(T)=\lim_{a \rightarrow \infty}
 \frac{\wt{Z}^{(m_1,m_2|n_1,n_2)}_{B,N}(aT)}{Z_{B,N}(aT)} \, ,
\label{c41}
\eeq
where $\wt{\mc{Z}}_{B,N}^{(m_1,m_2|n_1,n_2)}(T)$ and
$\wt{Z}_{B,N}^{(m_1,m_2|n_1,n_2)}(aT)$ represent 
canonical partition functions 
of the $BC_N$ type of anti-ferromagnetic
PF spin chain \eq{c39} and spin Calogero model (\ref{c40})
respectively.
The Hilbert space of    
$\wt{H}^{(m_1,m_2|n_1,n_2)}$ in (\ref{c40})
can be obtained as the closure of the linear subspace 
spanned by the wave functions which are quite 
similar to their ferromagnetic counterparts \eq{b18}.
More precisely, the state vectors associated 
with the Hilbert space of    
$\wt{H}^{(m_1,m_2|n_1,n_2)}$
are given by 
\beq
\tilde{\psi}_{\bf{r}}^{\bf{s}} 
\equiv 
\tilde{\psi}_{ r_1 ,\ldots, r_i, \ldots, r_j, \ldots , r_N}
^{s_1, \ldots, s_i, \ldots, s_j, \ldots, s_N}
=\tilde{\Lambda}^{(m_1,m_2|n_1,n_2)}
\left (\phi_{\bf{r}} ({\bf x})|{\bf s}\rangle \right) \, ,
\label{c42}
\eeq
where the completely symmetric projector
$\Lambda^{(m_1,m_2|n_1,n_2)}$ in \eq{b18}  
is replaced by  the completely antisymmetric 
projector $\tilde{\Lambda}^{(m_1,m_2|n_1,n_2)}$
related to the $BC_N$ type of Weyl algebra. 
Due to this change of the  projector, 
 the symmetry conditions \eq{b19} and \eq{b20} 
 of the wave functions are now modified as 
\beq
\tilde{\psi}_{ r_1 ,\ldots, r_i, \ldots, r_j, \ldots , r_N}
^{s_1, \ldots, s_i, \ldots, s_j, \ldots, s_N}
=(-1)^{{\alpha_{ij}(\mbf{s})}+1} \, 
\tilde{\psi}_{r_1 , \ldots, r_j, \ldots, r_i, \ldots, r_N}^
{s_1, \ldots, s_j, \ldots, s_i, \ldots , s_N} \, , 
\label{c43}
\eeq
and 
\beq
\tilde{\psi}_{r_1, \ldots, r_i, \ldots , r_N}^{s_1, \ldots, s_i, 
\ldots, s_N}=(-1)^{r_i+f(s_i)+1}~
\tilde{\psi}_{r_1, \ldots, r_i, \ldots , 
r_N}^{s_1, \ldots, s_i, \ldots, s_N}
\, ,
\label{c44}
\eeq
where $1\leq i<j \leq N$.
Due to these modified symmetry conditions, 
it is possible to  obtain a set of 
(non-orthonormal) basis vectors for 
the Hilbert space of $\wt{H}^{(m_1,m_2|n_1,n_2)}$ by 
 ordering the indices  $\mbf{s}$ and $\mbf{r}$ of the state vectors $\tilde{\psi}_{\bf{r}}^{\bf{s}}$ 
 in a suitable way. More precisely, 
the rule 1) described 
 in this section for ordering $\mbf{s}$ in 
 the ferromagnetic case remains unchanged in the present case,
while the rules 2) and 3) for ordering $\mbf{r}$ in 
 the ferromagnetic case 
are modified in the following way: 
  \vskip .2 cm 
$2'$)~If $s_i=s_j$ for $i<j$, then by using  \eq{c43}
the  components $r_i$ and $r_j$ within   
$\mbf{r} \equiv (r_1, r_2, \cdots ,r_N)$ are ordered   
such that $r_i\geqslant r_j+1-\pi(s_j)$. 
\vskip .11 cm
$3'$)~Due to the condition \eq{c44}, $r_i$ is taken as an odd
positive integer if $s_i^2 \equiv f(s_i)=0$ and $r_i$ is
taken as an even nonnegative integer if $s_i^2 \equiv f(s_i)=1$.
\vskip .11 cm
If one defines a partial ordering among
the above mentioned basis vectors as:
 $\tilde{\psi}_{\bf{r}}^{\bf{s}}>
 \tilde{\psi}_{\bf{r}'}^{ \bf{s}'} \, $, 
 for~$|\bf{r}|>|\bf{r}'|$,   
then  $\wt{H}^{(m_1,m_2|n_1,n_2)}$ in \eq{c40} would act   
as an upper triangular matrix on such partially ordered basis vectors. Consequently, 
all eigenvalues in the spectrum of $\wt{H}^{(m_1,m_2|n_1,n_2)}$   
are given by 
$E_{\mbf{r}}^{\mbf{s}}$ in \eq{b24}, 
where the indices $\mbf{r}$ and $\mbf{s}$  
are now ordered by using the set of rules $1),~2'),$ and $3')$.
Appropriately modifying the procedure described 
in this section 
for the ferromagnetic case,
in accordance with this new set of rules, 
we find that the grand  canonical partition function
of the $BC_N$ type of spin CS model \eq{c40} 
can be expressed as 
\beq
{\wt{\mbb{Z}}}^{(m_1,m_2|n_1,n_2)}_{B}
=\left\{ {\wt{\mbb{Z}}}^{(1,0|0,0)}_{B}  \right\}^{m_1}
\left\{ {\wt{\mbb{Z}}}^{(0,1|0,0)}_{B} \right\}^{m_2}
\left\{ {\wt{\mbb{Z}}}^{(0,0|1,0)}_{B}\right\}^{n_1}
\left\{ {\wt{\mbb{Z}}}^{(0,0|0,1)}_{B}\right\}^{n_2}
\, , 
\label{c45}
\eeq
where the anti-ferromagnetic basic modes appearing in the 
r.h.s. of the above equation are related to their 
ferromagnetic counterparts as 
\beq
{\wt{\mbb{Z}}}^{(1,0|0,0)}_{B}= {\mbb{Z}}^{(0,0|0,1)}_{B},~
{\wt{\mbb{Z}}}^{(0,1|0,0)}_{B}= {\mbb{Z}}^{(0,0|1,0)}_{B},~
{\wt{\mbb{Z}}}^{(0,0|1,0)}_{B}= {\mbb{Z}}^{(0,1|0,0)}_{B},~
{\wt{\mbb{Z}}}^{(0,0|0,1)}_{B}= {\mbb{Z}}^{(1,0|0,0)}_{B} .
\label{c46}
\eeq
Expanding the grand canonical partition function \eq{c45}
as a power series of the fugacity parameter,
we derive the corresponding 
canonical partition function as 
\beq
\wt{Z}^{(m_1,m_2|n_1,n_2)}_{B,N}(aT)
=\sum_{\stackrel{\sum\limits_{i=1}^{m_1} a_i+\sum\limits_{j=1}^{m_2} b_j
+\sum\limits_{k=1}^{n_1} c_k+\sum\limits_{l=1}^{n_2} d_l=N}
{a_i\geq 0,~ b_j\geq 0, ~c_k\geq 0, ~d_l\geq 0}}~
\frac{q^{\sum\limits_{i=1}^{m_1} a_i^2
+\sum\limits_{j=1}^{m_2}b_j(b_j-1)
+\sum\limits_{k=1}^{n_1} c_k}}
{\prod\limits_{i=1}^{m_1}(q^2)_{a_i}\prod\limits_{j=1}^{m_2}(q^2)_{b_j}\prod\limits_{k=1}^{n_1}(q^2)_{c_k}
\prod\limits_{l=1}^{n_2}(q^2)_{d_l}} \, .
\label{c47}
\eeq 
Substituting this expression of 
$\wt{Z}^{(m_1,m_2|n_1,n_2)}_{B,N}(aT)$, along with 
$Z_{B,N}(aT)$ given in \eq{b17}, to the relation \eq{c41},
we finally obtain the canonical partition function of the 
$BC_N$ type of anti-ferromagnetic PF chain \eq{c39} as
\beq
\wt{\mc{Z}}^{(m_1,m_2|n_1,n_2)}_{B,N}(q)
=\sum_{\stackrel{\sum\limits_{i=1}^{m_1} a_i+\sum\limits_{j=1}^{m_2} b_j
+\sum\limits_{k=1}^{n_1} c_k+\sum\limits_{l=1}^{n_2} d_l=N}
{a_i\geq 0,~ b_j\geq 0, ~c_k\geq 0, ~d_l\geq 0}}~
\frac{\left(q^2\right)_N \cdot 
q^{\sum\limits_{i=1}^{m_1} a_i^2
+\sum\limits_{j=1}^{m_2}b_j(b_j-1)
+\sum\limits_{k=1}^{n_1} c_k}}
{\prod\limits_{i=1}^{m_1}(q^2)_{a_i}\prod\limits_{j=1}^{m_2}(q^2)_{b_j}\prod\limits_{k=1}^{n_1}(q^2)_{c_k}
\prod\limits_{l=1}^{n_2}(q^2)_{d_l}} \, .
\label{c48}
\eeq 

In the next section,  we shall use 
the expressions \eq{c32} and  \eq{c48} for the partition
functions of the $BC_N$ type of PF spin chains 
for the purpose of constructing corresponding multivariate SRS polynomials. 

\noi \section{
SRS polynomials associated with $BC_N$ type of
PF spin chains}
\renewcommand{\theequation}{4.{\arabic{equation}}}
\setcounter{equation}{0}
\medskip
We have seen earlier that the homogeneous multivariate 
RS polynomial \eq{a3} is closely related to the 
partition function \eq{a2} 
of the $A_{N-1}$ type of \su{m} PF spin chain. 
Before starting the discussion on the $BC_N$ type of  homogeneous multivariate 
SRS polynomial, let us 
briefly review the connection between the 
partition function of the $A_{N-1}$ type of
\su{m|n} supersymmetric PF spin chain and 
the corresponding SRS polynomial \cite{HB00}.
The partition functions of the $A_{N-1}$ type of \su{m|n}
supersymmetric PF spins have been computed by using the 
freezing trick as~\cite{BUW99}
\beq
\mc{Z}^{(m|n)}_{A,N}(q)
=
\sum_{\stackrel{\sum\limits_{i=1}^{m} a_i+
\sum\limits_{j=1}^{n} b_j=N}
{a_i\geq 0,~ b_j\geq 0}}~
\frac{\left(q\right)_N \cdot q^{\sum\limits_{j=1}^{n} \frac{b_j(b_j-1)}{2}}}
{\prod\limits_{i=1}^{m}(q)_{a_i}\prod\limits_{j=1}^{n}(q)_{b_j}} \, .
\label{m1}
\eeq
It may be noted that, in the absence of the fermionic 
spin degrees of freedom (i.e., for the case $n=0$), 
$\mc{Z}^{(m|n)}_{A,N}(q)$
reduces to $\mc{Z}^{(m)}_{A,N}(q)$ in \eq{a2}.
Motivated by the form of the partition functions \eq{m1},  
the $A_{N-1}$ type of SRS polynomials 
have been defined as~\cite{HB00}
\beq
\mbb{H}_{A,N}^{(m|n)}(x,y;q)=
\sum_{\stackrel{\sum\limits_{i=1}^{m}a_i+\sum\limits_{j=1}^nb_j
=N}{a_i\geq 0,~b_j \geq 0}}(q)_N\cdot 
q^{\sum\limits_{j=1}^n\frac{b_j(b_j-1)}{2}}
\prod\limits_{i=1}^m
\frac{x_i^{a_i}}{(q)_{a_i}}\prod\limits_{j=1}^n
\frac{y_j^{b_j}}{(q)_{b_j}} \, ,
\label{m2}
\eeq
(along with $\mbb{H}_{A,0}^{(m|n)}(x,y;q)=1$), 
where $x\equiv x_1,x_2,\cdots,x_m$ and
$y\equiv y_1,y_2,\cdots,y_n$ represent
two different types of variables.
It is evident that 
the partition functions \eq{m1} can be obtained from the
SRS polynomials \eq{m2} as
$\mc{Z}_{A,N}^{(m|n)}(q)
=\mbb{H}_{A,N}^{(m|n)}(x=1,y=1;q)$.
Moreover, for the special case $n=0$, the SRS polynomial \eq{m2} reduces to its bosonic counterpart \eq{a3}.
By using the relation 
\beq
(q^{-1})_l=(-1)^l\, q^{-\frac{l(l+1)}{2}} \, (q)_l \, , 
\label{m3}
\eeq
the SRS polynomials \eq{m2} may be rewritten
as 
\beq
\mbb{H}_{A,N}^{(m|n)}(x,y;q)= 
\sum_{\stackrel{\sum\limits_{i=1}^{m}a_i+\sum\limits_{j=1}^nb_j
=N}{a_i\geq 0,~b_j \geq 0}}(q)_N \cdot 
\prod\limits_{i=1}^m
\frac{x_i^{a_i}}{(q)_{a_i}}\prod\limits_{j=1}^n
\frac{(-q^{-1}y_j)^{b_j}}{(q^{-1})_{b_j}} \, .
\label{m4}
\eeq
The above form of the SRS polynomials can be obtained from a 
power series expansion of 
the generating function given by~\cite{HB00} 
\beq
\mc{G}^{(m|n)}_A(x,y;q,t)=\frac{1}{\prod\limits_{i=1}^m(tx_i;q)_{\infty} \cdot
\prod\limits_{j=1}^n(-tq^{-1}y_j;q^{-1})_{\infty}} \, ,
\label{m5}
\eeq
where $(t;q)_0\equiv 1$ and  
$(t;q)_l \equiv (1-t)(1-qt)\cdots(1-q^{l-1}t)$ for $l>0$. 
Indeed, by using the identity \cite{An76} 
\beq
\frac{1}{(t:q)_{\infty}}=\sum\limits_{N=0}^{\infty}
\, \frac{t^N}{(q)_N} \, ,
\label{m6}
\eeq
it is easy to check that the generating function in \eq{m5} 
can be expanded as a 
power series of the parameter $t$ as 
\beq
\mc{G}^{(m|n)}_A(x,y;q,t)=\sum\limits_{N=0}^{\infty}
\frac{\mbb{H}_{A,N}^{(m|n)}(x,y;q)}{(q)_N} \, t^N \, . 
\label{m7}
\eeq

Inspired by the form of 
partition functions \eq{c32} of the $BC_N$ type of 
ferromagnetic PF chains with SAPSRO, 
we define  
$BC_N$ type of homogeneous multivariate
SRS polynomials of the first kind as 
\bea
&& \mbb{H}_{B,N}^{(m_1,m_2|n_1,n_2)}(x,\bar x,y,\bar y;q)\nn \\
&&=\sum_{\stackrel{\sum\limits_{i=1}^{m_1} a_i+\sum\limits_{j=1}^{m_2} b_j
+\sum\limits_{k=1}^{n_1} c_k+\sum\limits_{l=1}^{n_2} d_l=N}
{a_i\geq 0,~ b_j\geq 0, ~c_k\geq 0, ~d_l\geq 0}} 
\hskip -1.64 cm (q^2)_N \cdot 
 q^{\sum\limits_{j=1}^{m_2} b_j+\sum\limits_{k=1}^{n_1} c_k(c_k-1)+\sum\limits_{l=1}^{n_2} d_l^2} \cdot 
 \prod\limits_{i=1}^{m_1}\frac{x_i^{a_i}}{(q^2)_{a_i}}
\prod\limits_{j=1}^{m_2}\frac{(\bar{x}_j)^{b_j}}{(q^2)_{b_j}}
\prod\limits_{k=1}^{n_1}\frac{y_k^{c_k}}{(q^2)_{c_k}}
\prod\limits_{l=1}^{n_2}\frac{(\bar{y}_l)^{d_l}}{(q^2)_{d_l}}
 \, ,\nn \\
 &&~~~~
\label{m8}
\eea
and set
$\mbb{H}_{B,0}^{(m_1,m_2|n_1,n_2)}(x,\bar x,y,\bar y;q)=1$,
where $x\equiv x_1,x_2,\cdots,x_{m_1}$, 
$\bar x\equiv \bar x_1, \bar x_2,\cdots, \bar x_{m_2}$, 
$y\equiv y_1,y_2,\cdots,y_{n_1}$ and 
$\bar y\equiv \bar y_1, \bar y_2,\cdots, \bar y_{n_2}$
represent four different types of variables.
The partition functions 
in \eq{c32} can be obtained from these $BC_N$ type of 
SRS polynomials as 
\beq
\mc{Z}_{B,N}^{(m_1,m_2|n_1,n_2)}(q)
=\mbb{H}_{B,N}^{(m_1,m_2|n_1,n_2)}(x=1,\bar{x}=1,y=1,\bar{y}=1;q)
\, .
\label{m9}
\eeq
In the absence of the fermionic spin degrees of freedom, 
i.e., for the case  $n_1=n_2=0$, the $BC_N$ type of   
SRS polynomials \eq{m8} reduce to that type of   
RS polynomials of the form 
\beq
 \mbb{H}_{B,N}^{(m_1,m_2)}(x,\bar x;q)
=\sum_{\stackrel{\sum\limits_{i=1}^{m_1} a_i+\sum\limits_{j=1}^{m_2} b_j=N}
{a_i\geq 0,~ b_j\geq 0}} 
 (q^2)_N \cdot 
 q^{\sum\limits_{j=1}^{m_2} b_j} \cdot 
 \prod\limits_{i=1}^{m_1}\frac{x_i^{a_i}}{(q^2)_{a_i}}
\prod\limits_{j=1}^{m_2}\frac{(\bar{x}_j)^{b_j}}{(q^2)_{b_j}}
 \, .
\label{m10}
\eeq
Interestingly,
in another special case  
like $m_1=m$, $m_2=0$, $n_1=n$, $n_2=0$,
the $BC_N$ type of 
SRS polynomials \eq{m8} can be connected with the 
$A_{N-1}$ type of SRS polynomials \eq{m2} as 
\beq
\mbb{H}_{B,N}^{(m,0|n,0)}(x,y;q)
= \mbb{H}_{A,N}^{(m|n)}(x,y;q^2) \, . 
\label{m11}
\eeq
By using the relation \eq{m3}, we express
the $BC_N$ type of 
SRS polynomials \eq{m8} in a more compact form as 
\bea
&& \mbb{H}_{B,N}^{(m_1,m_2|n_1,n_2)}(x,\bar x,y,\bar y;q)\nn \\
&&=\sum_{\stackrel{\sum\limits_{i=1}^{m_1} a_i+\sum\limits_{j=1}^{m_2} b_j
+\sum\limits_{k=1}^{n_1} c_k+\sum\limits_{l=1}^{n_2} d_l=N}
{a_i\geq 0,~ b_j\geq 0, ~c_k\geq 0, ~d_l\geq 0}} 
\hskip -1.01 cm (q^2)_N \cdot 
 \prod\limits_{i=1}^{m_1}\frac{x_i^{a_i}}{(q^2)_{a_i}}
\prod\limits_{j=1}^{m_2}\frac{(q\bar{x}_j)^{b_j}}{(q^2)_{b_j}}
\prod\limits_{k=1}^{n_1}
\frac{(-q^{-2}y_k)^{c_k}}{(q^{-2})_{c_k}}
\prod\limits_{l=1}^{n_2}\frac{(-q^{-1}\bar{y}_l)^{d_l}}{(q^{-2})_{d_l}}
 \, .~~~~~~~~~
\label{m12}
\eea
Let us now define a generating function of the form 
\beq
\mc{G}_B^{(m_1,m_2|n_1,n_2)}(x,\bar{x},y,\bar{y};q,t)=\mc{G}_1^{(m_1)}(x;q,t) \cdot \mc{G}_2^{(m_2)}(\bar{x};q,t) \cdot 
\mc{G}_3^{(n_1)}(y;q,t) \cdot  
\mc{G}_4^{(n_2)}(\bar{y};q,t) \, ,
\label{m13}
\eeq
where
\begin{subequations}

\bea
&&\mc{G}_1^{(m_1)}(x;q,t)=\frac{1}{\prod\limits_{i=1}^{m_1}(tx_i;q^2)_{\infty}} \, , 
\label{m14a} \\
&&\mc{G}_2^{(m_2)}(\bar{x};q,t)=\frac{1}{\prod\limits_{j=1}^{m_2}(tq\bar{x}_j;q^2)_{\infty}} \, , 
\label{m14b} \\
&&\mc{G}_3^{(n_1)}(y;q,t)=\frac{1}{\prod\limits_{k=1}^{n_1}(-tq^{-2}y_k;q^{-2})_{\infty}} \, , \label{m14c} \\
&&\mc{G}_4^{(n_2)}(\bar{y};q,t)=\frac{1}{\prod\limits_{l=1}^{n_2}(-tq^{-1}\bar{y}_l;q^{-2})_{\infty}} \, . \label{m14d}
\eea
\label{m14}
\end{subequations}
Expanding all terms appearing in the r.h.s. of \Eq{m13} 
by using the identity \eq{m6} and subsequently using the
expression of the $BC_N$ type of SRS polynomials given in \eq{m12}, we obtain
\beq
\mc{G}_B^{(m_1,m_2|n_1,n_2)}(x,\bar{x},y,\bar{y};q,t)=\sum\limits_{N=0}^{\infty}\frac{
\mbb{H}_{B,N}^{(m_1,m_2|n_1,n_2)}(x,\bar x,y,\bar y;q)}
{(q^2)_N} \, t^N \, .
\label{m15}
\eeq
Thus $\mc{G}_B^{(m_1,m_2|n_1,n_2)}(x,\bar{x},y,\bar{y};q,t)$ 
in \eq{m13} represents the generating function of the  
$BC_N$ type of SRS polynomials.

We have already seen in \Eq{m11} that, in a particular case,
the $BC_N$ type of SRS polynomial 
can be expressed through the $A_{N-1}$ type of SRS polynomial.
For the purpose of exploring such connection between 
the $BC_N$ and $A_{N-1}$ types of SRS polynomials
in a general case,
we use Eqs.~\eq{m14a}, \eq{m14c} and \eq{m5} to find that  
\beq
\mc{G}_1^{(m_1)}(x;q,t) \cdot \mc{G}_3^{(n_1)}(y;q,t)
=\mc{G}_A^{(m_1|n_1)}(x,y;q^2,t).
\label{m16}
\eeq
Hence, by using the power series expansion \eq{m7}, we obtain
\beq
\mc{G}_1^{(m_1)}(x;q,t)\cdot \mc{G}_3^{(n_1)}(y;q,t)=\sum\limits_{N_1=0}^{\infty}
\frac{\mbb{H}_{A,N_1}^{(m_1|n_1)}(x,y;q^2)}{(q^2)_{N_1}
}\, t^{N_1}.
\label{m17}
\eeq
Next, by using Eqs.~\eq{m14b}, \eq{m14d} and \eq{m5},
we find that 
\beq
\mc{G}_2^{(m_2)}(\bar{x};q,t) \cdot \mc{G}_4^{(n_2)}(\bar{y};q,t)=\mc{G}_A^{(m_2|n_2)}(\tilde{x},\tilde{y};q^2,t),
\label{m18}
\eeq
where $\tilde{x} \equiv q\cdot\bar{x}$ and 
$\tilde{y} \equiv q \cdot \bar{y}$. Hence, by using \eq{m7},
we obtain
\beq
\mc{G}_2^{(m_2)}(\bar{x};q,t)\cdot \mc{G}_4^{(n_2)}(\bar{y};q,t)
=\sum\limits_{N_2=0}^{\infty}\frac{\mbb{H}_{A,N_2}^{(m_2|n_2)}(\tilde{x},\tilde{y};q^2)}{(q^2)_{N_2}}\, t^{N_2}.
\label{m19}
\eeq
Since $\mbb{H}_{A,N_2}^{(m_2|n_2)}(\tilde{x},\tilde{y};q^2)$
is a homogeneous polynomial of the variables
$\tilde{x},~\tilde{y}$
of order $N_2$,
the above equation can be rewritten as 
\beq
\mc{G}_2^{(m_2)}(\bar{x};q,t)\cdot
\mc{G}_4^{(n_2)}(\bar{y};q,t)
=\sum\limits_{N_2=0}^{\infty}\frac{q^{N_2}}{(q^2)_{N_2}} \,
\mbb{H}_{A,N_2}^{(m_2|n_2)}(\bar{x},\bar{y};q^2) \, 
t^{N_2}.
\label{m20}
\eeq
Inserting the series expansions \eq{m17} and \eq{m20} in \Eq{m13}, it is easy to find that 
\bea
&&\mc{G}_B^{(m_1,m_2|n_1,n_2)}(x,\bar{x},y,\bar{y};q,t) \nn \\ &&~~~~=\sum\limits_{N=0}^{\infty}t^N
\sum\limits_{N_1=0}^{N}
\frac{q^{N-N_1}}{(q^2)_{N_1}\cdot (q^2)_{N-N_1}}
\mbb{H}_{A,N_1}^{(m_1|n_1)}(x,y;q^2)
\cdot \mbb{H}_{A,N-N_1}^{(m_2|n_2)}(\bar{x},\bar{y};q^2) \, .
\label{m21}
\eea
Comparing the coefficients of $t^N$ in the  
r.h.s. of \eq{m15} and \eq{m21}, we finally obtain 
a relation between the 
$BC_N$ and $A_{N-1}$ types of SRS polynomials as 
\bea
&&\mbb{H}_{B,N}^{(m_1,m_2|n_1,n_2)}(x,\bar{x},y,\bar{y};q,t)
~~~~~~\nn \\
&&~~~~~~~=\sum\limits_{N_1=0}^N q^{N-N_1}
\binom{N}{N_1}_{q^2}
\mbb{H}_{A,N_1}^{(m_1|n_1)}(x,y;q^2)\cdot 
\mbb{H}_{A,N-N_1}^{(m_2|n_2)}(\bar{x},\bar{y};q^2) \, ,
\label{m22}
\eea
where the notation  
$\binom{N}{N_1}_{q^2} \equiv \qbinom{N}{N_1,N-N_1}{q^2}$
has been used. 

It may be noted that,
even though the $BC_N$ type of partition function  
given in \eq{c32} is rather complicated in form, 
it can be easily computed for arbitrary values of $N$
and for some small values of the discrete parameters $m_1,~m_2,~n_1$ and $n_2$.
In particular, by using \eq{c32}, it is easy to find that  
\beq
\mc{Z}_{B,N}^{(1,0|0,0)}(q)=1,~
\mc{Z}_{B,N}^{(0,1|0,0)}(q)=q^N, ~
\mc{Z}_{B,N}^{(0,0|1,0)}(q)=q^{N(N-1)},~
\mc{Z}_{B,N}^{(0,0|0,1)}(q)=q^{N^2}.
\label{m23}
\eeq
In this context it is interesting to ask whether
there exists some recursion
relations such that, by taking the 
 partition functions given in   
\eq{m23} as the initial
conditions, it is possible to compute
$\mc{Z}^{(m_1,m_2|n_1,n_2)}_{B,N}(q)$ 
for arbitrarily values of the discrete parameters 
$m_1,~m_2,~n_1,~n_2$ and $N$. To answer this question, 
we define a generating function as 
\beq
\mc{G}_B^{(m_1,m_2|n_1,n_2)}(q,t)
\equiv \mc{G}_B^{(m_1,m_2|n_1,n_2)}(x=1,\bar{x}=1,y=1,\bar{y}=1;q,t) \, .
\label{m24}
\eeq
Inserting $x=1$, $\bar{x}=1$, $y=1$, $\bar{y}=1$ in \eq{m15} 
and also using \eq{m9}, one can expand 
$\mc{G}_B^{(m_1,m_2|n_1,n_2)}(q,t)$ in a power series 
of $t$ as 
\beq
\mc{G}_B^{(m_1,m_2|n_1,n_2)}(q,t)=
\sum\limits_{N=0}^{\infty}
\frac{\mc{Z}_{B,N}^{(m_1,m_2|n_1,n_2)}(q)}{(q^2)_N} \, t^N \, ,
\label{m25}
\eeq
where it is assumed that 
$\mc{Z}_{B,0}^{(m_1,m_2|n_1,n_2)}(q)=1$. 
Therefore, $\mc{G}_B^{(m_1,m_2|n_1,n_2)}(q,t)$ may be considered as the generating function for the partition function
$\mc{Z}_{B,N}^{(m_1,m_2|n_1,n_2)}(q)$. Next, by using Eqs.~\eq{m13}, \eq{m14} and \eq{m24}, we find that 
\bea
\mc{G}_B^{(m_1,m_2|n_1,n_2)}(q,t)&& \nn \\
&&\hskip -2.83 cm =\frac{1}{\{(t;q^2)_{\infty}\}^{m_1} \cdot 
\{(tq;q^2)_{\infty}\}^{m_2} \cdot 
\{(-tq^2;q^{-2})_{\infty}\}^{n_1} \cdot 
\{(-tq^{-1};q^{-2})_{\infty}\}^{n_2}} \, . \nn  
\eea
Consequently, this generating function satisfies a
factorization relation given by 
\beq
\mc{G}_B^{(m_1+m_1^{\prime},m_2+m_2^{\prime}|n_1+n_1^{\prime},n_2+n_2^{\prime})}(q,t)=
\mc{G}_B^{(m_1,m_2|n_1,n_2)}(q,t) \cdot 
\mc{G}_B^{(m_1^{\prime},m_2^{\prime}|
n_1^{\prime},n_2^{\prime})}(q,t).
\label{m26}
\eeq
Expanding both sides of the above equation by using 
\eq{m25} and comparing the coefficients of $t^N$, 
we find that 
\beq
\mc{Z}_{B,N}^{(m_1+m_1^{\prime},m_2+m_2^{\prime}|n_1+n_1^{\prime},n_2+n_2^{\prime})}(q)=
\sum\limits_{N_1=0}^N \binom{N}{N_1}_{q^2}
\mc{Z}_{B,N-N_1}^{(m_1,m_2|n_1,n_2)}(q)  \cdot 
\mc{Z}_{B,N_1}^{(m_1^{\prime},m_2^{\prime}|n_1^{\prime},n_2^{\prime})} \, .
\label{m27}
\eeq
Appropriately choosing the values of discrete variables 
$m_1',~m_2',~n_1',~n_2'$ in 
\Eq{m27} and also using \Eq{m23}, we derive a set of recursion relations like 
\bea
&&\mc{Z}_{B,N}^{(m_1+1,m_2|n_1,n_2)}(q)=\sum\limits_{N_1=0}^N {\binom{N}{N_1}_{q^2}}\cdot \mc{Z}_{B,N-N_1}^{(m_1,m_2|n_1,n_2)}(q)
\, , \nonumber\\
&&\mc{Z}_{B,N}^{(m_1,m_2+1|n_1,n_2)}(q)=\sum\limits_{N_1=0}^N
{\binom{N}{N_1}_{q^2}}\cdot q^{N_1}
\cdot \mc{Z}_{B,N-N_1}^{(m_1,m_2|n_1,n_2)}
(q) \, , \nonumber\\
&&\mc{Z}_{B,N}^{(m_1,m_2|n_1+1,n_2)}(q)=
\sum\limits_{N_1=0}^N{\binom{N}{N_1}_{q^2}} \cdot  
q^{N_1(N_1-1)} \cdot 
\mc{Z}_{B,N-N_1}^{(m_1,m_2|n_1,n_2)}(q) \, ,\nonumber\\
&&\mc{Z}_{B,N}^{(m_1,m_2|n_1,n_2+1)}(q)=\sum\limits_{N_1=0}^N
{\binom{N}{N_1}_{q^2}}\cdot q^{N_1^2} \cdot   
\mc{Z}_{B,N-N_1}^{(m_1,m_2|n_1,n_2)}(q) \, . 
\label{m28}
\eea
By using this set of recursion relations and also using the 
initial conditions in \eq{m23}, it is possible to compute  
$\mc{Z}^{(m_1,m_2|n_1,n_2)}_{B,N}(q)$
for arbitrarily values of the discrete parameters 
$m_1,~m_2,~n_1,~n_2$ and $N$. Furthermore, 
it is easy to check that,  all initial 
conditions appearing in \eq{m23} can also be derived from the 
recursion relations \eq{m28} by using only one initial 
condition given by
\beq
\mc{Z}^{(0,0|0,0)}_{B,N}(q)=\delta_{N,0} \, . 
\label{m29}
\eeq

Next, we define  
$BC_N$ type of homogeneous multivariate
SRS polynomials of the second kind as 
\bea
&& \wt{\mbb{H}}_{B,N}^{(m_1,m_2|n_1,n_2)}
(x,\bar x,y,\bar y;q)\nn \\
&&=\sum_{\stackrel{\sum\limits_{i=1}^{m_1} a_i+\sum\limits_{j=1}^{m_2} b_j
+\sum\limits_{k=1}^{n_1} c_k+\sum\limits_{l=1}^{n_2} d_l=N}
{a_i\geq 0,~ b_j\geq 0, ~c_k\geq 0, ~d_l\geq 0}} 
\hskip -1.64 cm (q^2)_N \cdot 
 q^{\sum\limits_{i=1}^{m_1}a_i^2+ \sum\limits_{j=1}^{m_2}b_j(b_j-1)
+\sum\limits_{k=1}^{n_1}c_k} \cdot 
 \prod\limits_{i=1}^{m_1}\frac{x_i^{a_i}}{(q^2)_{a_i}}
\prod\limits_{j=1}^{m_2}\frac{(\bar{x}_j)^{b_j}}{(q^2)_{b_j}}
\prod\limits_{k=1}^{n_1}\frac{y_k^{c_k}}{(q^2)_{c_k}}
\prod\limits_{l=1}^{n_2}\frac{(\bar{y}_l)^{d_l}}{(q^2)_{d_l}}
 \, ,\nn \\
 &&~~~~
\label{m30}
\eea
which is related to the 
partition function \eq{c48} associated with the $BC_N$ type of anti-ferromagnetic PF spin chain as 
\beq
\wt{\mc{Z}}_{B,N}^{(m_1,m_2|n_1,n_2)}(q)
=\wt{\mbb{H}}_{B,N}^{(m_1,m_2|n_1,n_2)}(x=1,\bar{x}=1,y=1,\bar{y}=1;q)
\, .
\label{m31}
\eeq
Comparing \eq{m30} with \eq{m8}, we find that
$BC_N$ type of SRS polynomials of the first kind
and the second kind are related as
\beq
\wt{\mbb{H}}_{B,N}^{(m_1,m_2|n_1,n_2)}(x,\bar x,y,\bar y;q)
=\mbb{H}_{B,N}^{(n_2,n_1|m_2,m_1)}(\bar y,y,\bar x,x;q) \ .
\label{m32}
\eeq
It may be noted that, by using Eqs.~\eq{m32} and
\eq{m22}, one can easily 
derive a relation between $BC_N$ type of SRS polynomials of the second kind and $A_{N-1}$ type of SRS polynomials. 
Let us now try to find out the   
generating function for 
$\wt{\mbb{H}}_{B,N}^{(m_1,m_2|n_1,n_2)}
(x,\bar x,y,\bar y;q)$,  
which would satisfy the relation 
\beq
\wt{\mc{G}}_B^{(m_1,m_2|n_1,n_2)}(x,\bar{x},y,\bar{y};q,t)=\sum\limits_{N=0}^{\infty}\frac{
\wt{\mbb{H}}_{B,N}^{(m_1,m_2|n_1,n_2)}(x,\bar x,y,\bar y;q)}
{(q^2)_N} \, t^N \, .
\label{m33}
\eeq
Using Eqs.~\eq{m15}, \eq{m32} and \eq{m33},
it is easy to find that 
\beq
\wt{\mc{G}}_B^{(m_1,m_2|n_1,n_2)}(x,\bar{x},y,\bar{y};q,t)=
\mc{G}_B^{(n_2,n_1|m_2,m_1)}(\bar{y},y,\bar{x},x;q,t) \, . 
\label{m34}
\eeq
By using the above relation along with \eq{m13}, we 
get an expression for this generating function as  
\beq
\wt{\mc{G}}_B^{(m_1,m_2|n_1,n_2)}(x,\bar{x},y,\bar{y};q,t)=\mc{G}_1^{(n_2)}(\bar{y};q,t) \cdot \mc{G}_2^{(n_1)}(y;q,t) \cdot 
\mc{G}_3^{(m_2)}(\bar{x};q,t) \cdot  
\mc{G}_4^{(m_1)}(x;q,t) \, ,
\label{m35}
\eeq
where the factors appearing in the r.h.s. can be obtained
from \Eq{m14}. Using the $x=\bar{x}=y=\bar{y}=1$
limit of this generating function and 
following a procedure similar to the ferromagnetic
case, it can be shown that 
$\wt{\mc{Z}}_{B,N}^{(m_1,m_2|n_1,n_2)}(q)$ 
satisfies a relation
exactly of the form \eq{m27}.
 For some small values of the discrete parameters $m_1,~m_2,~n_1$, $n_2$, and for arbitrary values 
of $N$,  \Eq{c48} yields  
\beq
\wt{\mc{Z}}_{B,N}^{(1,0|0,0)}(q)=q^{N^2},~
\mc{Z}_{B,N}^{(0,1|0,0)}(q)=q^{N(N-1)}, ~
\mc{Z}_{B,N}^{(0,0|1,0)}(q)=q^{N},~
\mc{Z}_{B,N}^{(0,0|0,1)}(q)=1.
\label{m36}
\eeq
Using these partition functions 
and an equation of the form
\eq{m27} corresponding to the anti-ferromagnetic case, 
we derive a set of recursion relations like
\bea
&&\wt{\mc{Z}}_{B,N}^{(m_1+1,m_2|n_1,n_2)}(q)=\sum\limits_{N_1=0}^N {\binom{N}{N_1}_{q^2}}
\cdot q^{N_1^2} \cdot 
\wt{\mc{Z}}_{B,N-N_1}^{(m_1,m_2|n_1,n_2)}(q) \, , \nonumber\\
&&\wt{\mc{Z}}_{B,N}^{(m_1,m_2+1|n_1,n_2)}(q)=\sum\limits_{N_1=0}^N {\binom{N}{N_1}_{q^2}}
\cdot q^{N_1(N_1-1)} \cdot 
\wt{\mc{Z}}_{B,N-N_1}^{(m_1,m_2|n_1,n_2)}(q) \, , \nonumber\\
&&\wt{\mc{Z}}_{B,N}^{(m_1,m_2|n_1+1,n_2)}(q)=
\sum\limits_{N_1=0}^N {\binom{N}{N_1}_{q^2}}
\cdot q^{N_1} \cdot 
\wt{\mc{Z}}_{B,N-N_1}^{(m_1,m_2|n_1,n_2)}(q) \, , \nonumber\\
&&\wt{\mc{Z}}_{B,N}^{(m_1,m_2|n_1,n_2+1)}(q)=\sum\limits_{N_1=0}^N {\binom{N}{N_1}_{q^2}} \cdot 
\wt{\mc{Z}}_{B,N-N_1}^{(m_1,m_2|n_1,n_2)}(q) \, . \nonumber\\
\label{m37}
\eea
By using this set of recursion relations
and 
the initial conditions given in \eq{m36}
(or, alternatively, a single initial condition
of the form \eq{m29}), in principle
it is possible to compute  
$\wt{\mc{Z}}^{(m_1,m_2|n_1,n_2)}_{B,N}(q)$
for arbitrarily values of the discrete parameters 
$m_1,~m_2,~n_1,~n_2$ and $N$. Indeed by using 
the symbolic software package  Mathematica
we have seen that, in comparison to the 
direct use of the expressions \eq{c32} and \eq{c48},
it is much more efficient to use the  
set of corresponding recursion relations  
\eq{m28}  and \eq{m37} 
to obtain explicit  
forms of the ferromagnetic and anti-ferromagnetic
partition functions
as some polynomials of the variable $q$.
Hence, the 
set of recursion relations  \eq{m28} and \eq{m37}
might be useful in studying various spectral properties
like level density distribution and nearest neighbour 
spacing distribution
for the $BC_N$ type of ferromagnetic and anti-ferromagnetic
PF chains.

\bigskip

\noi \section{Conclusions}
\renewcommand{\theequation}{6.{\arabic{equation}}}
\setcounter{equation}{0}
\medskip
 
Here we derive the canonical partition functions 
of the $BC_N$ type of PF spin chains with 
SAPSRO by employing the freezing trick via the indirect
approach,
and subsequently construct the related  
$BC_N$ type of homogeneous 
multivariate SRS polynomials.
More precisely, 
we  compute the grand canonical 
partition functions of the $BC_N$ type of ferromagnetic 
as well as anti-ferromagnetic spin
Calogero models with SAPSRO, and expand those   
grand canonical 
partition functions as some power series 
of the fugacity parameter 
to obtain the corresponding canonical 
partition functions. Applying the freezing trick, 
subsequently we  
 derive novel expressions for the canonical  
partition functions of the related $BC_N$ type of
PF spin chains.
Inspired by the form of such partition functions, 
we  define $BC_N$ type of homogeneous 
multivariate SRS polynomials and also 
find out the corresponding
generating functions. 
Using these generating
functions, we   
show that the $BC_N$ type of SRS polynomials
can be expressed as some bilinear 
combinations of the $A_{N-1}$ type of SRS polynomials.

It is worth noting that,  
the grand canonical 
partition functions of the $BC_N$ type of spin
Calogero models with SAPSRO are expressed
as some simple products of only $BC_N$ types
of basic modes in Eqs.~\eq{c25} and \eq{c45}. 
Such expressions of the grand canonical
partition functions play an important role 
in our construction of the   
$BC_N$ type of SRS polynomials. 
Even though the  
grand canonical partition functions of the $BC_N$ type of 
spin Calogero models with SASRO have been computed 
earlier~\cite{BFGR09},  it was found that 
both $BC_N$ and $A_{N-1}$ types of basic modes appear in the expressions of such grand canonical partition functions.
Comparing our results with this earlier work, we find 
novel relations like (3.36a,b),
which connect the basic modes 
of the  $A_{N-1}$ and $BC_N$ types of
grand canonical partition functions
and also lead to interesting
 $q$-identities of the form \eq{c37} and \eq{c38}.
 
In this paper, we also 
derive a set of recursion relations \eq{m28} and \eq{m37} 
for the partition functions of the $BC_N$ type of 
PF spin chains involving different numbers of 
lattice sites and internal degrees of freedom. 
In this context it may be noted that, another type of
recursion relations, 
involving different numbers of 
lattice sites and fixed values of the internal degrees 
of freedom,
have been computed earlier for the partition functions of the $A_{N-1}$ type of PF (supersymmetric PF) spin chains  
and the corresponding RS (SRS) polynomials 
\cite{Hi95npb, Hi95, KKN97,HB00}. The later type of recursion relations play a key role in expressing 
the spectra of the 
$A_{N-1}$ type of PF spin chains and their supersymmetric
generalizations through the motifs and  in constructing the 
related one-dimensional vertex models.
It can be shown that 
 the $BC_N$ type of SRS polynomials considered in the present  
 paper also satisfy the later type of 
recursion relations, 
involving different values of $N$ and 
fixed values of the internal parameters   
$m_1,~m_2,~n_1$ and $n_2$. As a result, the spectra of  
$BC_N$ type of PF spin chains with SAPSRO can be described
by some motif like objects similar to the case of 
$A_{N-1}$ type of spin chains. Furthermore,
it is possible to construct
  some one-dimensional classical
vertex models whose energy functions would generate 
the complete spectra of these $BC_N$ type of  PF spin chains.
We plan to describe such exciting developments on the 
 $BC_N$ type of  PF spin chains and related SRS polynomials
 in a forthcoming publication.

\bigskip 
\noi {\bf Acknowledgements}
\medskip

We thank Artemio Gonz\'alez-L\'opez and Federico Finkel 
for some discussions. This work is supported by the TPAES 
project of Theory Division, 
Saha Institute of Nuclear Physics, India.

\newpage 

\bibliographystyle{model1a-num-names}
\bibliography{cmprefs}

\end{document}